\documentclass[aps,pra,twocolumn,reprint, superscriptaddress]{revtex4-2}

\usepackage[svgnames]{xcolor}
\usepackage{amsmath}
\usepackage{microtype}
\usepackage{amssymb}
\usepackage{amsmath}
\usepackage{braket}
\usepackage{graphicx}
\usepackage{booktabs}
\usepackage{tikz}
\usetikzlibrary{quantikz}
\usepackage{bm}
\usepackage{dsfont}
\usepackage{hyperref}

\hypersetup{colorlinks,linkcolor={DarkBlue},citecolor={DarkBlue},urlcolor={DarkBlue}}

\let\originalleft\left
\let\originalright\right
\renewcommand{\left}{\mathopen{}\mathclose\bgroup\originalleft}
\renewcommand{\right}{\aftergroup\egroup\originalright}

\newcommand{\id}{\mathds{1}}
\DeclareMathOperator\erfc{erfc}

\begin{document}

\title{Fault-tolerant quantum computation with static linear optics}

\author{Ilan Tzitrin}
\thanks{These authors contributed equally.}
\affiliation{Xanadu, Toronto, ON, M5G 2C8, Canada}
\email{ilan@xanadu.ai}
\affiliation{Department of Physics, University of Toronto, Toronto ON, M5S 1A7, Canada}
\author{Takaya Matsuura}
\thanks{These authors contributed equally.}
\affiliation{Xanadu, Toronto, ON, M5G 2C8, Canada}
\affiliation{Department of Applied Physics, Graduate School of Engineering, The University of Tokyo, 7–3–1 Hongo, Bunkyo-ku, Tokyo 113–8656, Japan}
\author{Rafael N.~Alexander}
\thanks{These authors contributed equally.}
\affiliation{Xanadu, Toronto, ON, M5G 2C8, Canada}
\affiliation{Centre for Quantum Computation and Communication Technology, School of Science, RMIT University, Melbourne, VIC 3000, Australia}
\affiliation{Center for Quantum Information and Control, University of New Mexico, Albuquerque, NM 87131, USA}
\author{Guillaume Dauphinais}
\affiliation{Xanadu, Toronto, ON, M5G 2C8, Canada}
\author{J.~Eli Bourassa}
\affiliation{Xanadu, Toronto, ON, M5G 2C8, Canada}
\affiliation{Department of Physics, University of Toronto, Toronto ON, M5S 1A7, Canada}
\author{Krishna K.~Sabapathy}
\affiliation{Xanadu, Toronto, ON, M5G 2C8, Canada}
\author{Nicolas C.~Menicucci}
\affiliation{Xanadu, Toronto, ON, M5G 2C8, Canada}
\affiliation{Centre for Quantum Computation and Communication Technology, School of Science, RMIT University, Melbourne, VIC 3000, Australia}
\author{Ish Dhand}
\affiliation{Xanadu, Toronto, ON, M5G 2C8, Canada}

\date{\today}

\begin{abstract}
The scalability of photonic implementations of fault-tolerant quantum computing based on Gottesman-Kitaev-Preskill (GKP) qubits is injured by the requirements of inline squeezing and reconfigurability of the linear optical network. In this work we propose a topologically error-corrected architecture that does away with these elements at no cost---in fact, at an advantage---to state preparation overheads. Our computer consists of three modules: a 2D array of probabilistic sources of GKP
states; a depth-four circuit of static beamsplitters, phase shifters, and short delay lines; and a 2D array of homodyne detectors. The symmetry of our proposed circuit allows us to combine the effects of finite squeezing and uniform photon loss within the noise model, resulting in more comprehensive threshold estimates. These jumps over both architectural and analytical hurdles considerably expedite the construction of a photonic quantum computer.
\end{abstract}

\maketitle
\textit{Introduction.---}The photonic quantum computing paradigm is well-placed to handle the long-term obstacles inherent to engineering scalable quantum computers. The promise of this technology is enabled by room-temperature functionality, manufacturability, tolerance to photon loss, and the potential for long-range networking. In this approach, the need for robust and stable optical quantum information is met by combining bosonic codes known as Gottesman-Kitaev-Preskill (GKP) qubits~\cite{gottesman2001encoding} with qubit quantum error correcting codes implemented through measurement-based quantum computation (MBQC)~\cite{raussendorf2001one, menicucci2006universal, raussendorf2006}, in a hybrid continuous-variable (CV) and discrete-variable (DV) architecture~\cite{wu2020quantum, bourassa2020, fukui2020temporal, zhu2019hypercubic, larsen2021fault}. However, the current best architectures of this type still have critical challenges: inline squeezing in circuit or measurement-based implementations of CZ gates introduce noise~\cite{yoshikawa2008demonstration,larsen2020deterministic, larsen2021fault,bourassa2020}; the requirement of deterministic GKP sources leads to onerous multiplexing costs;
and the need for rapid reconfiguration in the linear optics networks is a substantial burden on integrated chips~\cite{collins2013integrated}. All of these elements furthermore increase the number of optical components seen in each photon's journey, thereby compounding loss---the most harmful imperfection in a photonic quantum computer.

Here we show how to entangle the outputs of probabilistic sources of GKP qubits into fault-tolerant resource states for MBQC without requiring either inline squeezing or reconfigurable linear optics. Our architecture produces a three-dimensional macronodal lattice structure~\cite{menicucci2008one,flammia2009optical, menicucci2011temporal,yokoyama2013ultra, chen2014experimental, wang2014weaving, asavanant2019generation, larsen2019deterministic, wu2020quantum, fukui2020temporal, larsen2021fault} in one temporal and two spatial dimensions where each site consists of four modes. The advantage of this approach is that the generation circuit consists only of single-mode sources, a depth-4 static circuit of balanced beamsplitters, half-time-step delay lines, and homodyne detectors. The generated resource state can be used equivalently to the CV/DV hybridized Raussendorf-Harrington-Goyal (RHG) cluster state~\cite{bourassa2020,raussendorf2005long, raussendorf2006, raussendorf2007topological}, although the process is generalizable to other qubit codes. Furthermore, both finite squeezing noise and uniform photon loss throughout the beamsplitter network are equivalent to local Gaussian noise before each detector due to the symmetry of the generation circuit. 

We calculate logical error rates of the outer (qubit) code for different levels of finite squeezing and photon loss, over a range of failure probabilities of GKP state generation. In the event that a source fails to produce a GKP state, we assume it produces a squeezed vacuum state. We find, for example, that at 15~dB of squeezing and no loss, our architecture can tolerate GKP failure rates of more than $50\%$, reducing by a significant factor the size of the per-node state preparation modules and multiplexers in Ref.~\cite{bourassa2020}. In addition, under the condition of deterministic GKP state generation, we find a squeezing threshold of $\sim 10$ dB, lower than that found in Ref.~\cite{bourassa2020}, despite the latter neglecting the noise from inline squeezing within the CZ gates. Finally, we show the simple trade-off between tolerable finite squeezing noise and uniform photon loss rates for a given GKP failure rate. 

\textit{Background.---}
Qubits are encoded into optical bosonic modes by the GKP encoding, with ideal logical 0 and 1 codewords defined as
\begin{equation}
    \ket{\bar{\mu}} = \sum_{n} \Ket{(2n+\mu)\sqrt{\pi}}_q,\,\mu=0,1,
\end{equation}
where $\ket{\cdot}_q$ is a position eigenstate.
Throughout the paper, single-mode states within the GKP code space are indicated with an overbar.
Given a single-mode squeezer $S(\xi) \coloneqq \exp(-i (\ln\xi) (\hat{q}\hat{p}+\hat{p}\hat{q}))$, the states needed in our scheme are a momentum eigenstate $\ket{0}_{p}$, the sensor state $\ket{\varnothing} = S(\sqrt{2})\ket{\bar{+}}$, and a magic state such as
$S(\sqrt{2})\ket{+\bar{T}}$, where $\ket{+\bar{T}}\coloneqq \frac{1}{\sqrt{2}}(\ket{\bar{0}} + e^{\frac{\pi}{4} i}\ket{\bar{1}})$, the last of which is required to implement non-Clifford operations.

The effects of finite squeezing are modelled by the application of an additive Gaussian bosonic channel on the ideal $\ket{0}_p$ and $\ket{\varnothing}$ states\,\cite{menicucci2014fault}: 
\begin{equation}
    \tilde{\mathcal{N}}[\epsilon](\rho)\coloneqq \iint \frac{dr\, ds}{2\pi\epsilon}\, e^{-\frac{r^2}{2\epsilon}-\frac{s^2}{2\epsilon}} X(r)Z(s) \rho Z^{\dagger}(s)X^{\dagger}(r),
    \label{eq:additive_gaussian}
\end{equation}
where $X(r)\coloneqq \exp(-ir \hat{p})$ and $Z(s)\coloneqq \exp(i s \hat{q})$ are displacements along the position and momentum phase-space directions, respectively.

The 50:50 beamsplitter is defined as $B_{jk} \coloneqq e^{-i\pi (\hat{q}_{j} \hat{p}_{k}- \hat{p}_{j} \hat{q}_{k})/4} = B_{kj}^{\dag}$, and depicted by an arrow from mode $j$ to $k$ \cite{walshe2020}. The phase shifter is defined as $R(\theta)\coloneqq e^{i\theta \hat{n}}$, with  $R(\pi/2)$ corresponding to a Fourier transform in phase space, which implements a GKP Hadamard gate. Homodyne detectors measure linear combinations of the quadrature operators, with $\hat{q}$, $\hat{p}$, and $\hat{q} + \hat{p}$ measurements implementing GKP Pauli $Z$, $X$, and $Y$ measurements, respectively. The single-mode squeezed vacuum state is given by $ S(\xi)\ket{0}$ with $\ln{\xi}\rightarrow \pm\infty$ being $\ket{0}_{p(q)}$. The CV CZ gate is defined as $C\!Z_{ jk}\coloneqq e^{i  \hat{q}_j\hat{q}_k}$, and the CV CX gate as $C\!X_{ jk}=e^{-i  \hat{q}_j\hat{p}_k}$. These implement GKP CZ and CX gates, respectively. In this article, we differentiate $C\!X_{ jk}$ from $C\!X_{ jk}^{\dagger}$ by using a solid vs. open circle on the control mode $j$, respectively. 
Finally, GKP Pauli X and Z operators are realized by displacements of any odd-integer-multiple of $\sqrt{\pi}$ in the $q$ and $p$ quadratures, respectively.

\begin{figure}
    \centering
    \includegraphics[width=\linewidth]{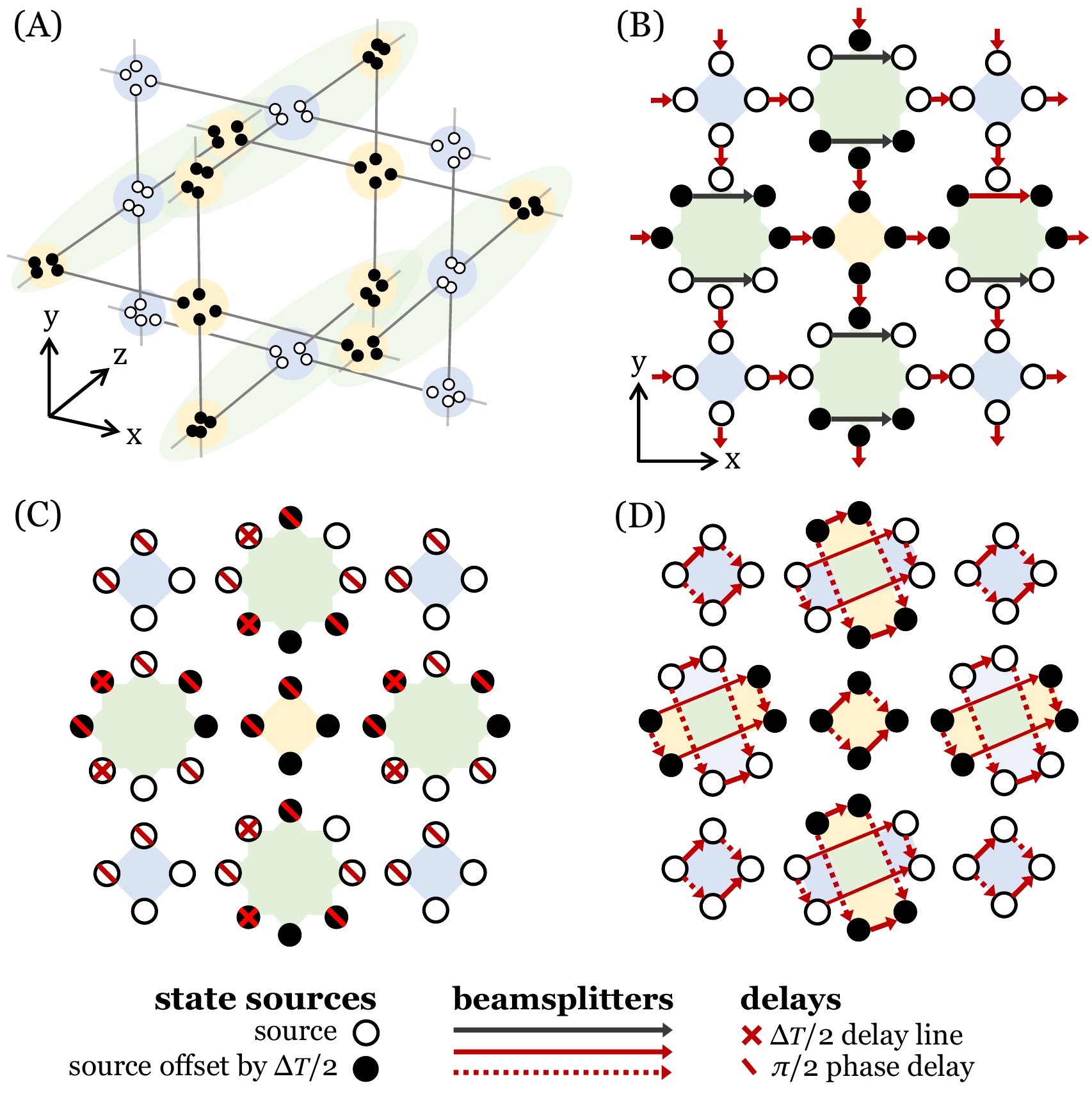}
    \caption{\textbf{(A)} shows the primal unit cell of the 3D hybrid pair cluster state and (B)--(C) shows the steps for generating it. (B)--(D) are presented as cross-sections of waveguide layers stacked in the Z direction, which coincides with the direction of propagation of light through the waveguides. The 3D lattice exists in two spatial (X, Y) dimensions and one temporal dimension. The latter is divided into discrete time-bins of width $\Delta T$.  Colors are included for the relationship between sources and the final state. \textbf{(B)} Waveguide arrangement at the first layer, with each node receiving an input from a source in every $\Delta T$-wide time bin. The time bins for the solid nodes are offset by $\Delta T/2$, relative to the hollow nodes. 50:50 beamsplitters are applied between pairs of modes as indicated by the arrows, and these generate entangled pairs (see Eq.~\eqref{eq:beamsplitter_cz}). 
    The beamsplitters indicated by black arrows create entangled pairs that will connect the state in the Z direction.
    In \textbf{(C)}, Xs indicate the application of a $\Delta T/2$ time-delay line, while slashes indicate the application of a $\pi/2$ phase delay.
    In \textbf{(D)}, the state is connected into the macronode cluster state by the application of four additional beamsplitters between the four modes that make up each macronode. Dotted beamsplitters are applied after solid ones. 
    Notice that the time signature of certain nodes changes due to the time-delay lines.
    }
    \label{fig:layer}
\end{figure}

\textit{3D hybrid macronode architecture.---}Ref.~\cite{bourassa2020} proposed a constant-depth generation circuit for the RHG lattice state compatible with probabilistic GKP state sources.
Though simple in theory, this proposal remains experimentally challenging because it requires inline squeezing (present in the CZ gates) and time-varying circuits (different gate arrangements between even and odd time steps). Both these problems can be circumvented by substituting the RHG lattice target state with a computationally equivalent \emph{macronode cluster state}~\cite{menicucci2008one,flammia2009optical, menicucci2011temporal,yokoyama2013ultra, chen2014experimental, wang2014weaving, asavanant2019generation, larsen2019deterministic, wu2020quantum, fukui2020temporal, larsen2021fault}, where each node has several modes that undergo multi-mode measurement.

The basic building block of our scheme is a type of two-mode entangled state, which can be produced by first generating a pair of modes---either being GKP or momentum squeezed vacuum---and sending these through a 50:50 beamsplitter. Though the constituent modes are coupled only by a beamsplitter, the resulting pairs are equivalent to two-mode cluster states, as is made apparent by the following identities \cite{walshe2020}:
\begin{equation}
B_{jk} \coloneqq
\begin{quantikz}[row sep=0.2cm, column sep=0.4cm]
\lstick{$k$} & \qw & \ghost{S^{\dagger}(\sqrt{2})}\qw \\
 \lstick{$j$} & \qw \arrow[u] &\ghost{S^{\dagger}(\sqrt{2})} \qw
\end{quantikz}
= \begin{quantikz}[row sep=0.2cm, column sep=0.4cm]
    & \octrl{1} & \gate{\, S(\sqrt{2}) \,} & \targ{} & \qw \\
    & \targ{} & \gate{S^{\dagger}(\sqrt{2})} & \ctrl{-1} & \qw
\end{quantikz},
\label{eq:beamsplitter_cnot}
\end{equation}
\begin{align}
    &C\!X_{ jk}^{\dagger}\ket{\varnothing}_j\ket{\varnothing}_k = \ket{\varnothing}_j\ket{\varnothing}_k\label{eq:cxprop1},\\
    &C\!X_{ jk}^{\dagger}\ket{\varphi}_j\ket{0}_{p_k} = \ket{\varphi}_j\ket{0}_{p_k}\label{eq:cxprop2}, \\
    &C\!X_{ jk}^{\dagger}\ket{0}_{q_j}\ket{\varphi}_k = \ket{0}_{q_j}\ket{\varphi}_k,
\end{align}
where $\ket{\varphi}$ can be an arbitrary state. From these identities, we get 
\begin{equation}
\begin{tikzpicture}
\node[scale=0.9]{
\begin{quantikz}[row sep=0.2cm, column sep=0.5cm]
    \lstick{$S(\sqrt{2})\ket{\psi}$} &\gate{R(-\tfrac{\pi}{2})}& \qw & \gate{R(\tfrac{\pi}{2})} & \qw \\
    \lstick{$S(\sqrt{2})\ket{\phi}$} & \qw & \qw \arrow[u] & \ghost{R(\tfrac{\pi}{2})}\qw & \qw
\end{quantikz}
\quad = 
\begin{quantikz}[row sep=0.2cm, column sep=0.5cm]
    \lstick{$\ket{\psi}$} & \ctrl{1} & \ghost{R(\tfrac{\pi}{2})} \qw \\
    \lstick{$\ket{\phi}$} & \control{} & \ghost{R(\tfrac{\pi}{2})}\qw
\end{quantikz},
};
\end{tikzpicture}
\label{eq:beamsplitter_cz}
\end{equation}
provided that both $\ket{\psi}$ and $\ket{\phi}$ are $\ket{\bar{+}}$, or at least one of the states $\ket{\psi},\, \ket{\phi}$ be in $\ket{0}_p$. Even if one has access only to either $\ket{\varnothing}$ or $\ket{0}_p$ at random, one always obtains an entangled state that functions as a unit of a hybrid CV-GKP qubit cluster. Magic states can be inserted into our architecture by letting $\ket{\psi}$ or $\ket{\phi}$ be a magic state such as $\ket{+\bar{T}}$, while letting the other be $\ket{0}_p$. 

We require that these entangled pairs be arranged in a 3D configuration, shown in FIG.~\ref{fig:layer}(A). To achieve this, we begin with a 2D array of sources that emit $\ket{\varnothing}$ with probability $1-p_{\rm swap}$ and momentum-squeezed states with probability $p_{\rm swap}$ at regular intervals.
Following Ref.~\cite{bourassa2020}, we assume the desired probabilities $p_{\rm swap}$ may arise from multiplexing multiple GBS sources for each effective source~\footnote{Technically, only GKP code states were considered previously\,\cite{su2019,Tzitrin2019}. However, because the state $\ket{\varnothing}$ differs from these only by a squeezing operation, and such operations are assumed ``free'' on the output of the GBS states, the requirements for state generation can be easily accounted for.}. We require that each source produces an input mode every time step of length $\Delta T$, though the timing of half of the sources is off-set by $\Delta T/2$ according to its location in the 2D layout in FIG.~\ref{fig:layer}(B). The beamsplitters, delay lines, and phase delays in FIG.~\ref{fig:layer}(B) and (C) produce the required arrangement of pair states in (2+1) dimensions. 

To create a fully connected 3D resource state, we apply four 50:50 beamsplitters within each macronode, as shown in FIG.~\ref{fig:layer}(D), analogously to the so-called quad-rail lattice construction~\cite{menicucci2011temporal,wang2014,alexander2016flexible}. A detailed graphical representation of the resulting state is given in Appendix~\ref{sec:graph}. Each mode is subsequently sent to a homodyne detector.

\begin{figure*}[ht]
    \centering
    \includegraphics[width=0.98\linewidth]{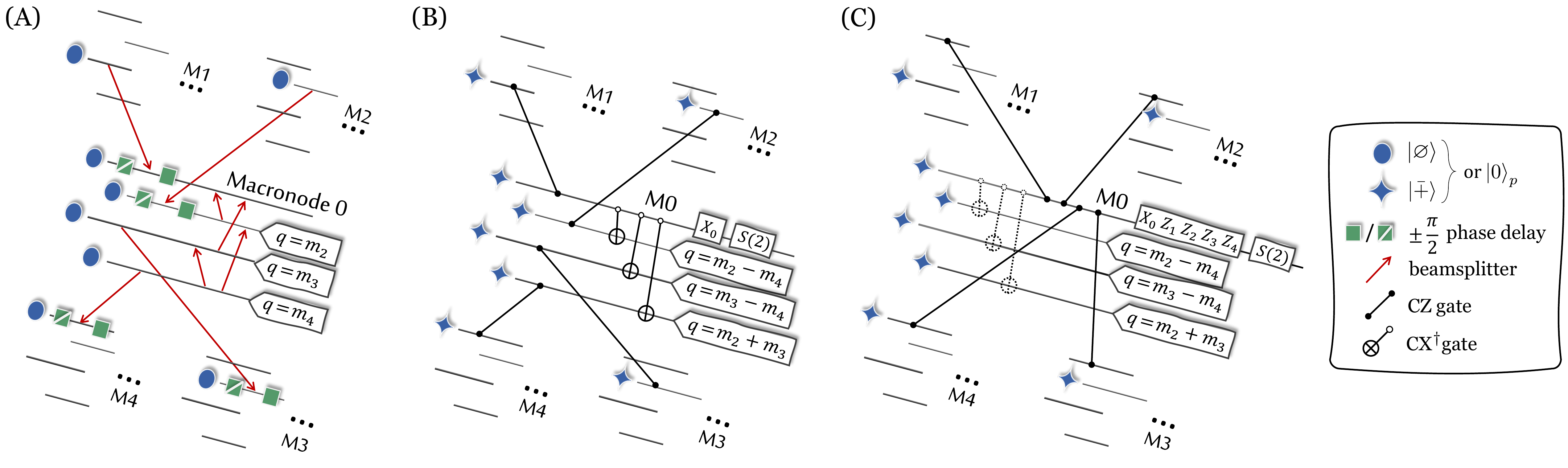}
        \caption{
            \textbf{(A)} Circuit representation of the beamsplitter network associated with a single macronode, 0, in the case where the central mode is the top wire. Also shown is the connectivity, by beamsplitters, to neighboring macronodes. See the legend for circuit conventions.
            The final four beamsplitters correspond to those in FIG.~\ref{fig:layer}(D).   
            \textbf{(B)} Equivalent circuit to (A), which follows from application of identities~\eqref{eq:beamsplitter_cz} and \eqref{eq:four_beamsplitter_identified}.
            $X_0$ denotes displacement $X\left((m_2+m_3-m_4)/2\right)$ as shown in the circuit~\eqref{eq:four_beamsplitter_identified}. $S$ is the squeezing gate defined in the main text, whose effect is to rescale the homodyne outcomes. \textbf{(C)} Equivalent circuit to (B), which follows from circuit identities that migrate CZ gates toward the measurements. The commuted $C\!X^{\dagger}$ gates are depicted with a dotted line because they act trivially on the circuit input (see the main text). The displacements $Z_{1, \ldots, 4}$ depend on the measurement outcomes of satellite modes in the neighboring macronodes according to the rules in App.~\ref{subsec:processing}.
            }  
    \label{fig:circuit_identity}
\end{figure*}

\textit{Equivalence to the canonical hybrid cluster state.---}
We refer to the hybrid RHG cluster state proposed in Ref.~\cite{bourassa2020} as the \emph{canonical RHG lattice state} since there is one mode per node and its generation involves CZ gates~\cite{menicucci2011graphical}. The state produced by the circuit in FIG.~\ref{fig:layer} is a macronode version of this state. We consider the case of always measuring three modes, referred to as satellite modes, within a macronode in the $\hat{q}$ basis. The remaining mode, referred to as central mode, then forms the canonical RHG lattice state. Here, we prove this through circuit identities. 

To simplify the description of the post-measurement state, we have the central mode in each macronode be chosen from wires whose inputs are prepared in GKP states, whenever possible. Representing the state generation and measurement via a quantum circuit, we can further simplify to the case where the central mode is taken to be the top wire shown in FIG.~\ref{fig:circuit_identity}\,(A), as other cases can be made equivalent to this one by permuting the measurement bases at the end~\cite{alexander2016flexible}. Using Eqs.~\eqref{eq:beamsplitter_cnot} and \eqref{eq:beamsplitter_cz}, we can replace the beamsplitters with CX$^{(\dagger)}$ gates and squeezers. 
At the measurement side, applying the commutation relations between gates $X(a)$, $S(\xi)$, and $C\!X_{ jk}$, as well as the identities $\bra{m}_{q}X(a) = \bra{m-a}_{q}$, $\bra{m}_{q}S(\xi) \propto \bra{m/\xi}_{q}$, and $\bra{m}_{q_1} C\!X_{ 1k} = \bra{m}_{q_1} X_k(m)$ for homodyne measurements, we obtain the equivalent circuit shown in FIG.~\ref{fig:circuit_identity}(B). 
Next, we commute all the CZ gates across the CX gates, towards the measurements, using the relation $C\!X_{ 1k}^{\dagger}C\!Z_{ jk}=C\!Z_{ jk}C\!Z_{j1}C\!X_{ 1k}^{\dagger}$. This generates additional CZ gates, but those with support on satellite modes can be replaced with displacements by the identity $\bra{m}_{q_k} C\!Z_{ jk} = \bra{m}_{q_k} Z_j(m)$. These changes are shown in FIG.~\ref{fig:circuit_identity}\,(C).
A detailed step-by-step derivation is given in Appendix\,\ref{sec:circuit_identity}.

Since we assumed the central mode to be an encoded GKP state $\ket{\bar{\psi}}_{1}$---either a plus state or a magic state---if the macronode contains at least one GKP state, then, by using Eq.~\eqref{eq:cxprop2} and
\begin{equation}
    C\!X_{ jk}^{\dagger}\ket{\bar{\psi}}_{j}\ket{\bar{+}}_{k}=\ket{\bar{\psi}}_{j}\ket{\bar{+}}_{k},
\end{equation}
we can remove the $C\!X_{ 1j}^{\dagger}$ $(j\in\{2,3,4\})$ gates that act at the beginning of the circuit in FIG.~\ref{fig:circuit_identity}~(C) \footnote{The number of magic states must be at most one per macronode.}. Therefore, the satellite modes are decoupled from the entanglement structure, and the state supported on just the central modes of each macronode is identical to the hybrid RHG lattice considered in Ref.~\cite{bourassa2020}---up to squeezing ($S(2)$) and displacement operators ($X_0$ and $Z_1$--$Z_4$), whose effect can be eliminated in post-processing. This treatment has thus far ignored the effects of finite squeezing and photon loss, so we now turn to their inclusion. 

\textit{Noise model.---}Any single-mode Gaussian bosonic channel $\mathcal{E}$ that preserves the phase-space mean of the vacuum state satisfies  
\begin{equation}
    B_{jk}\bigl(\mathcal{E}_{j}\otimes\mathcal{E}_{k}(\cdot)\bigr)B_{jk}^{\dagger} = \mathcal{E}_{j}\otimes\mathcal{E}_{k}\bigl(B_{jk}(\cdot)B_{jk}^{\dagger}\bigr).
    \label{eq:channelcomm}
\end{equation}
Furthermore, if $\mathcal{E}$ is also isotropic with respect to phase-space quadratures, then
\begin{equation}
    R(\theta) \bigl(\mathcal{E}(\cdot)\bigr)R^{\dagger}(\theta) = \mathcal{E}\bigl(R(\theta)(\cdot)R^{\dagger}(\theta)\bigr).
    \label{eq:channelcomm2}
\end{equation}
From these identities, it follows that uniform photon loss occurring just before the beamsplitter layers in  FIG.~\ref{fig:layer}~(B--D) can be combined and commuted across to act immediately before the layer of homodyne detectors in FIG.~\ref{fig:circuit_identity}~(A). Let $\eta$ denote the total transmission coefficient of the accumulated losses acting before each detector. By rescaling the homodyne outcomes by $1/\sqrt{\eta}$, the accumulated loss channel can be replaced with a Gaussian random displacement channel with variance $\sigma_{\text{loss}}^{2}=\frac{1-\eta}{2\eta}$~\cite{fukui2020all}. Finite squeezing noise, modelled as a Gaussian random displacement with $\sigma^{2}_{\text{fin.\,sq.}}$  as shown in Eq.~\eqref{eq:additive_gaussian} acting on the raw outputs of the sources, can similarly be commuted across all the optical elements so that it acts before the homodyne detectors. The combined effects of both losses and finite squeezing noise lead to homodyne outcomes with an uncertainty drawn from a normal distribution with variance $\sigma^{2}_{\text{total}} = \sigma^{2}_{\text{fin.\,sq.}} + \sigma^{2}_{\text{loss}}$.

Now that the photon loss and finite squeezing noise are accounted for as Gaussian random noise in the measurement data, one is free to apply the reduction to the canonical RHG lattice state described above.  However, reinterpreting this noisy measurement data to undo the conditional displacements (a.k.a. \emph{byproduct operators}) on the central mode in FIG.~\ref{fig:circuit_identity}~(C) will further distort the homodyne outcome of the central mode. More details of the noise model are given in App.~\ref{subsec:error_analysis}.

\begin{figure}[t]
    \centering
    \includegraphics[width=\linewidth]{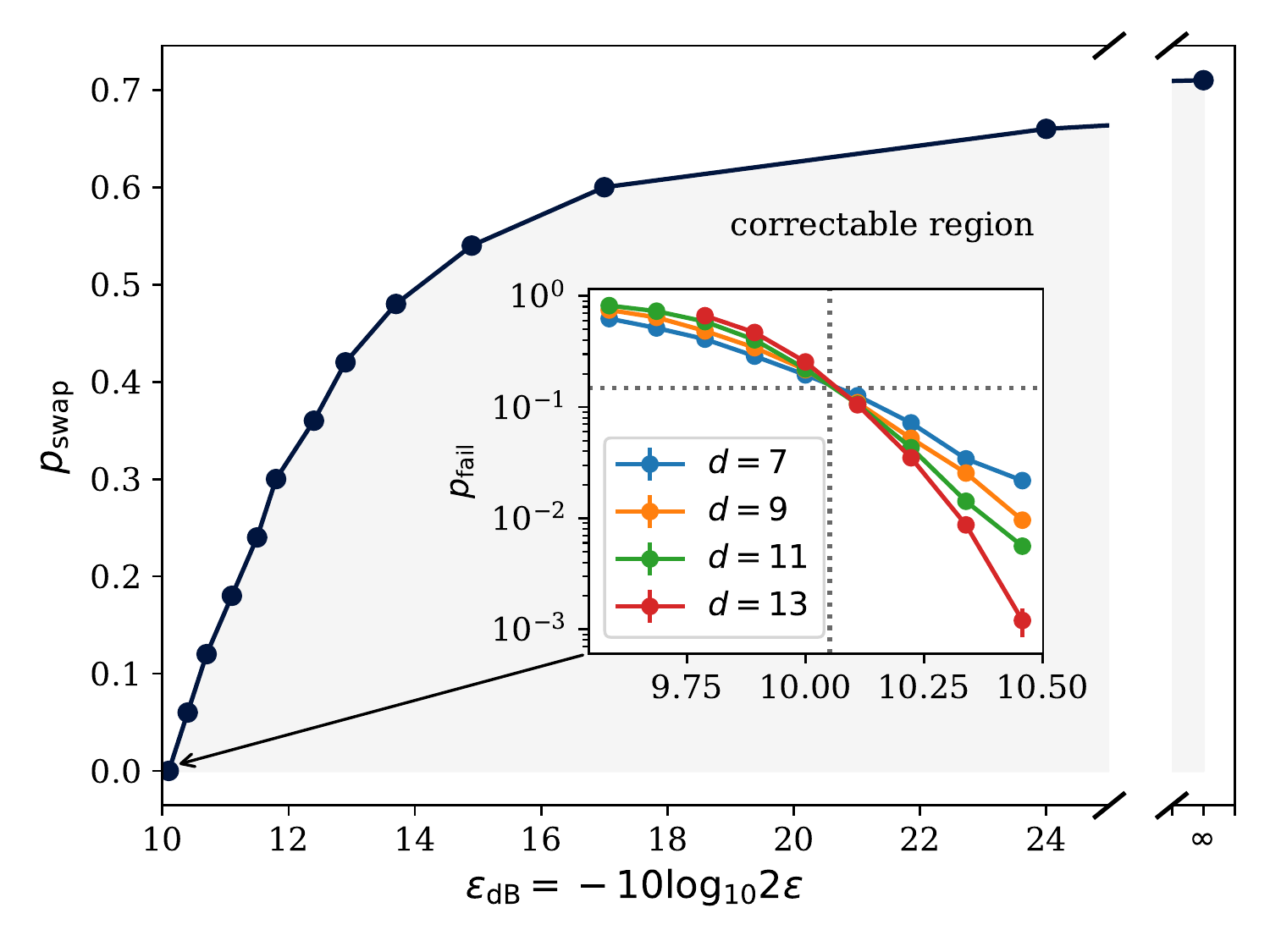}
    \caption{Swap-out probability $p_\text{swap}$ over physical error threshold $\epsilon$ for the passive architecture. $\epsilon$ combines the effects of finite squeezing (parameter $\sigma^2_{\text{fin.\,sq.}}$) and uniform loss (parameter $\sigma^2_{\text{loss}} = \frac{1- \eta}{2\eta}$ for transmissivity $\eta$
    ) through $\epsilon = \sigma^2_{\text{fin.\,sq.}} + \sigma^2_{\text{loss}}$. Each navy blue point reflects a Monte Carlo threshold search and fit for a given $p_\text{swap}$. A minimum-weight-perfect-matching decoder is used with matching graph weights assigned according to estimated qubit-level error probabilities on each node, as in App.~\ref{subsec:error_analysis} and \ref{sec:app_threshold}. We find the error threshold starts at 10.1~dB in the all-GKP case and tends to infinity as the swap-out probability approaches 0.71. This makes our passive and static architecture significantly more tolerant to swap-outs compared to the blueprint~\cite{bourassa2020}. \textbf{Inset}: Logical failure probability over noise parameter $\epsilon$ for an all-GKP macronode RHG code of varying odd distances. We also find an improvement to the no-swap-out threshold compared to~\cite{bourassa2020}.
    }
\label{fig:swap_tol}
\end{figure}

\textit{Threshold calculations.---} We find the correctable region for our macronode resource state through Monte Carlo simulations, where each trial comprises of three steps: simulating the complete macronode RHG lattice prepared in FIG.~\ref{fig:layer}, reducing it to the canonical lattice, and performing error correction on the reduced lattice.  

The noisy homodyne outcomes of the macronode lattice are generated by first sampling the (ideal) quadratures, applying the entangling gates, and then using them as the means of a normal distribution with a covariance matrix $\sigma_{\rm total}^2 \id$. This model corresponds to uniform Gaussian pre-detection noise. Following the above-described reduction procedure, noise on the central modes originates from both the generation circuit and from the byproduct operators. App.~\ref{subsec:processing} describes the post-processing rules, and App.~\ref{subsec:error_analysis} describes the propagating noise applied to the outcomes on the central modes. Conditional qubit-level error probabilities can then be estimated and used for decoding of the higher-level code. We proceed as in Ref.~\cite{bourassa2020} through minimum-weight-perfect matching~\cite{Edmonds1965}. The details of the simulation are presented in  App.~\ref{sec:app_threshold}.  

Thresholds calculated for various swap-out probabilities $p_\text{swap}$ are shown in FIG.~\ref{fig:swap_tol}. In the case where all modes are in GKP states (inset of FIG.~\ref{fig:swap_tol}), we find a threshold of 10.1 dB. With the additional restriction that every macronode has exactly one GKP state, the threshold becomes 13.6 dB (see FIG.~\ref{fig:p_fixed_threshold}). There is a marked improvement in the swap-out tolerance of our passive architecture: it is approximately 71\%, compared with the $\sim 24\%$ figure determined in~\cite{bourassa2020}. These values---occurring at the limit of infinite squeezing and no loss---invite comparison because of the unchanged decoder between the architectures. We leave open the possibly of a better decoder further increasing our swap-out tolerance. On the other hand, comparison with~\cite{larsen2021fault} is more tenuous, since there the authors assume an all-GKP encoding followed by rounds of explicit GKP error correction on the states.

We offer two main reasons for the observed improvement. First, as was described in Ref.~\cite{bourassa2020}, swapping a GKP mode with a momentum squeezed state introduces noise correlated among its neighbors. Our analysis (see App.~\ref{sec:circuit_identity}) reveals that reduced lattice will have an effective momentum squeezed state only if all four modes in the pre-reduced macronode were swapped out. Thus, the redundancy in the macronode lattice results in a greater tolerance to swap outs. Second, byproduct operators conditioned on the measurements of neighboring GKP states are binned, and thus do not propagate Gaussian noise; in fact, every additional GKP state present in a given macronode provides an additional degree of local GKP error correction.  

\textit{Discussion.---}
Previous work showed how quantum error correction (in the form of a topologically protected cluster state) can be used for photonic quantum computation with probabilistic sources of GKP qubits, provided that the available squeezing is sufficiently high~\cite{bourassa2020, larsen2021fault}. However, that work also required both inline squeezing and time-varying beamsplitters, both of which are difficult to implement at the required noise levels. By using a static linear-optical circuit to generate a macronode lattice, the present architecture circumvents these obstacles, making it feasible to implement topological error correction at noise levels compatible with Ref.~\cite{bourassa2020}.

Our architecture does away with experimentally demanding CZ gates, which were assumed ideal in the analysis of Ref.~\cite{bourassa2020} and shown to substantially degrade the quality of the state in Ref.~\cite{larsen2021fault}. The culprit is the requirement of inline squeezing~\cite{braunstein2005squeezing, bourassa2021fast}. In essence, our construction avoids this by migrating all squeezing---with the aid of circuit identities---either to the input of the circuit, where it can be absorbed into the state preparation \cite{Sabapathy2019, su2019, Quesada2019, Tzitrin2019}, or to the output, where it manifests as classical processing of the homodyne measurement outcomes. Furthermore, by doubling the number of modes at sites with connectivity in the Z direction, our scheme eliminates the need for reconfigurability of optical elements in the cluster state generation circuit. The only remaining reconfigurable components are in the multiplexed sources of individual GKP states (where switches are required) and in the local oscillator phase of each homodyne detector. 

By exploiting the symmetry of our resource generation circuit, we show that both uniform loss and finite squeezing effects can be consolidated into a combined Gaussian noise associated with each detector \footnote{Indeed, every mode does not go through the same elements in the generation circuit and would generally not experience the same amount of loss. We leave the analysis of non-uniform loss to future work, but we remark that we are always free to add more noise through beamsplitters coupled to the environment in order to make losses at all modes equal}. This remarkably simple model reveals that finite squeezing noise and photon losses can be treated on the same footing, allowing us to go further in tackling experimentally consequential noise than prior work~\cite{bourassa2020, larsen2021fault}. 

Circuit identities reveal the built-in redundancy supplied by satellite modes of our resource state, arising from the permutation symmetry of the generation circuit \cite{alexander2016flexible}. Having multiple GKP states per macronode is tantamount to additional rounds of GKP error correction, which keeps the threshold around 10 dB in all-GKP case, even better than that of Ref.~\cite{bourassa2020}. But bestowing a macronode with even just one GKP state means the encoded state at each site still behaves like a GKP state, leading to significantly higher tolerance to swap-outs. At 15 dB---the current highest reported level of optical squeezing observed (for a squeezed vacuum in bulk optics)~\cite{vahlbruch2016detection}---our architecture can afford more than half of its GKP states to be replaced by momentum-squeezed states, compared to just $\sim 14 \%$ in~\cite{bourassa2020}. This means that the increase in the number of modes of the cluster is balanced by a corresponding decrease in the number of required probabilistic state sources in every node, which significantly relaxes the multiplexing requirements. Taken together, our results thus substantially facilitate the realization of a fault-tolerant and scalable photonic quantum computer.

\begin{acknowledgments}
The authors are grateful for discussions with John Sipe and Michael Vasmer. I.T. is supported by Mitacs through the Mitacs Accelerate program grant. J.E.B. is supported through an Ontario Graduate Scholarship, and by Mitacs through the Mitacs Accelerate program grant. The authors thank SOSCIP and SciNet~\cite{Loken_2010} for their computational resources. Computations have been performed on the Niagara supercomputer~\cite{10.1145/3332186.3332195} at the SciNet-SOSCIP HPC Consortium. SciNet is funded by: the Canada Foundation for Innovation under the auspices of Compute Canada; the Government of Ontario; Ontario Research Fund - Research Excellence; and the University of Toronto. SOSCIP is funded by the Federal Economic Development Agency of Southern Ontario, the Province of Ontario, IBM Canada Ltd., Ontario Centres of Excellence, Mitacs and 15 Ontario academic member institutions.
\end{acknowledgments}

\vfill

\pagebreak
\appendix

\section{Entanglement structure}\label{sec:graph}
\begin{figure*}
    \centering
    \includegraphics[width=\linewidth]{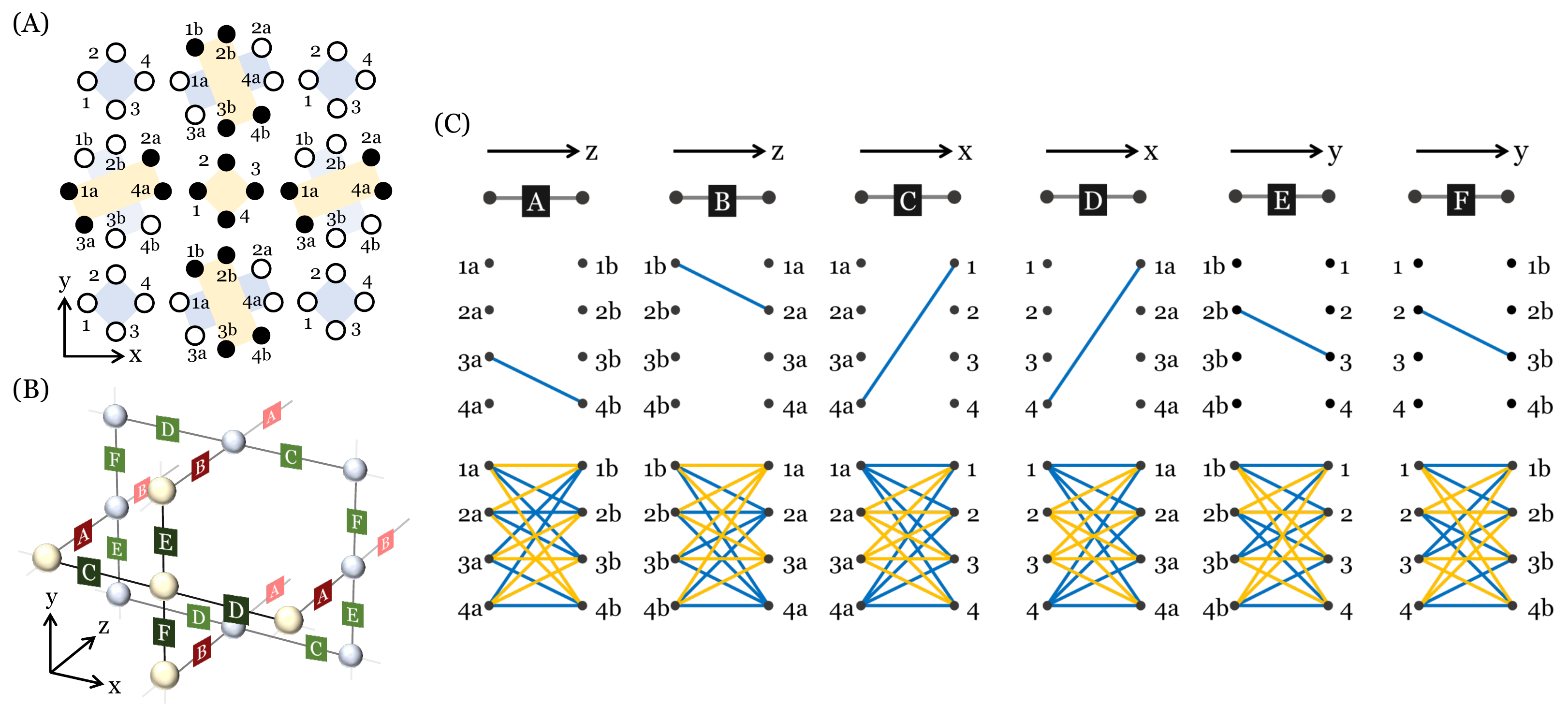}
    \caption{Graph of the hybrid macronode 3D cluster state. 
       \textbf{(A)} The 2D mode layout, with $\Delta T/2$ off-set modes consistent with stages C and D in FIG.~\ref{fig:layer}---i.e., modes at solid nodes are offset in time by $\Delta T/2$ relative to the modes at hollow nodes. Each macronode consists of four modes, labelled 1-4. \textbf{(B)} The three-dimensional arrangement of the four-mode macronodes is shown. For clarity, the plaques with letters A to F have a color corresponding to a given layer, with lighter colors in deeper layers (green colors for X and Y  direction and red for Z). We omit five modes from the unit cell, corresponding to the back face of the cube. The connectivity in the XY plane is identical to the front face.  \textbf{(C)} Macronode graph edges for each bond in B). The top six configurations correspond to weight-1 CZ gates, connecting pairs of modes as in FIG.~\ref{fig:layer} A.  The bottom six configurations correspond to weight $\pm \frac{1}{4}$ CZ gates (the sign is indicated by blue/yellow edge coloring, respectively), showing connectivity of the modes after the stage D in FIG.~\ref{fig:layer}.}
    \label{fig:fulllattice}
\end{figure*}

Here we provide a detailed description of the relationship between the generation circuit and the entanglement structure of the resulting state. After the state generation stages shown in FIG.~\ref{fig:layer}~(C) and (D), the array of modes is as shown in FIG.~\ref{fig:fulllattice}(A).

Recall that the solid nodes in FIG.~\ref{fig:layer} indicate that these lattice sites are present in temporal modes offset by $\Delta T/2$ relative to those in hollow nodes. The grouping of modes into macronodes is indicated by yellow (blue) squares and rectangles, also indicating that those macronodes are offset (not offset) by $\Delta T/2$. When the resource is constructed up to the point between stages C and D in FIG.~\ref{fig:layer}, it is equivalent to a projected entangled pair state (PEPS)~\cite{verstraete2004renormalization} for the CV/DV RHG cluster state, as shown in FIG.~\ref{fig:layer} (A). The precise identification of waveguide modes with graph nodes is given in FIG.~\ref{fig:fulllattice}.

In the main text, we describe a four-to-one reduction of modes for each macronode that corresponds to applying projectors equivalent to doing four beamsplitters and three homodyne measurements. This description provides a very natural way to understand how to implement computation on the reduced canonical state, but we stress that, in principle, other measurements could be performed after the beamsplitters (resulting in operations more general than what can be achieved on the canonical lattice). For completeness, we also present the four-layer graph for the state after beamsplitters but before any homodyne measurements, as shown in FIG.~\ref{fig:fulllattice} (C). Blue/yellow edge colorings are consistent with the plus/minus sign on the state's real-valued adjacency matrix that arises from the graphical calculus for Gaussian pure states in the case of all modes initially being squeezed states~\cite{menicucci2011graphical}. 

\section{Reduction to the RHG lattice} \label{sec:circuit_identity}
In this section, we give a step-by-step description of the equivalence between the macronode cluster state we actually generate, and the canonical RHG lattice state considered in Ref.~\cite{bourassa2020}. This complements the description in FIG.~\ref{fig:circuit_identity}\,A--C in the main text.  
  
\subsection{Restructuring the four-body measurement}
Here we recast the beamsplitter-based four-body measurement circuit described in FIG.~\ref{fig:circuit_identity}\,A into a circuit involving CX gates. The starting point is the following circuit:
\begin{equation}
\begin{quantikz}[row sep=0.3cm, column sep=0.6cm]
    & \qw  & \qw & \qw  & \qw & \qw \rstick{\text{``Central''}}  \\
    & \qw \arrow[u]   & \qw \vqw{1} \arrow[u]  & \qw & \measuretab{q = m_2} \rstick[wires=3]{\text{``Satellite''}}   \\
    & \qw & \qw  & \qw \vqw{1} \arrow[u] & \measuretab{q=m_3}  \\
    & \qw \arrow[u] & \qw & \qw & \measuretab{q = m_4}   \\
\end{quantikz}
\label{eq:four_beamsplitters}
\end{equation}
We name three modes measured in $\hat{q}$-basis the ``satellite modes'' and the remaining (topmost) mode the ``central mode.''
The last beamsplitter between the second and fourth modes in the circuit~\eqref{eq:four_beamsplitters} is equivalent to post-processing the measurement data  
\begin{align}
m_2\mapsto \frac{m_2-m_4}{\sqrt{2}}\quad \text{and}\quad m_4\mapsto \frac{m_2+m_4}{\sqrt{2}},
\end{align}
since 
\begin{align}
    {}_{q_2}\!\bra{m_2}{}_{q_4}\!\bra{m_4}B_{4 2}={}_{q_2}\!\left\langle\frac{m_2-m_4}{\sqrt{2}}\right\vert {}_{q_4}\!\left\langle\frac{m_2+m_4}{\sqrt{2}}\right\vert.
\end{align}    
Therefore, we remove it from the circuit, but take it into account by altering the measurement outcomes.

Next, applying the circuit identities~\eqref{eq:beamsplitter_cnot} to the circuit~\eqref{eq:four_beamsplitters} and using 
\begin{align}
    {}_{q_j}\!\!\bra{m}C\!X_{ jk} = {}_{q_j}\!\!\bra{m}X_k(m),
\end{align}
we find that the following circuit is equivalent to the circuit~\eqref{eq:four_beamsplitters}.  

\begin{equation}
\begin{tikzpicture}
\node[scale=0.75]{
\begin{quantikz}[row sep=0.3cm, column sep=0.4cm]
    & \octrl{1} & \gate{S(\sqrt{2})} & \gate{X\bigl(\frac{m_2-m_4}{\sqrt{2}}\bigr)} & \octrl{2} & \gate{S(\sqrt{2})} & \gate{X(m_3)} & \qw \\ 
    & \targ{} & \gate{S^{\dagger}(\sqrt{2})} & \measuretab{q = \frac{m_2-m_4}{\sqrt{2}}} \vcw{-1}  \\
    & \octrl{1} & \gate{S(\sqrt{2})} & \gate{X\bigl(\frac{m_2+m_4}{\sqrt{2}}\bigr)} & \targ{} & \gate{S^{\dagger}(\sqrt{2})} & \measuretab{q=m_3} \vcw{-2}  \\
    & \targ{} & \gate{S^{\dagger}(\sqrt{2})} & \measuretab{q = \frac{m_2+m_4}{\sqrt{2}}} \vcw{-1}   \\
\end{quantikz}
};
\end{tikzpicture}
\label{eq:replacing_beamsplitter}
\end{equation}

\noindent We can account for the action of squeezing operator on a homodyne outcome using 
\begin{align}
\!\bra{m}_qS(\xi)^{\dag} \propto \!\bra{\xi m}_q.
\end{align}
Furthermore, the squeezing operators can be conjugated through the displacement operators using 
\begin{align}
    S(\xi)^{\dagger} X(a) S(\xi) = X(\xi^{-1}a),
\end{align} 
and through the CX${}^{\dag}$ gates using $[S_i(s)S_j(s), e^{i q_i  p_j}]=0$ to obtain
\begin{equation}
\begin{tikzpicture}
\node[scale=0.8]{
\begin{quantikz}[row sep=0.3cm, column sep=0.4cm]
    & \octrl{1} & \gate{X\bigl(\frac{m_2-m_4}{2}\bigr)} & \octrl{2} & \gate{X(\frac{m_3}{2})} & \gate{S(2)} & \qw \\
    & \targ{} & \measuretab{q = m_2-m_4} \\
    & \octrl{1} & \gate{X\bigl(\frac{m_2+m_4}{2}\bigr)} & \targ{} & \qw & \measuretab{q=m_3} \\
    & \targ{} & \measuretab{q = m_2+m_4} \\
\end{quantikz}.
};
\end{tikzpicture}
\end{equation}

\noindent Next, we conjugate the displacement operators across the CX${}^{\dagger}$ gate using 
\begin{align}
   C\!X_{ jk}^{\dagger} X_j(a)X_k(b) = X_j(a)X_k(b-a)C\!X_{ jk}^{\dagger},
\end{align}
and push displacement operators into the measurements, to obtain 
\begin{equation}\label{eq:id1}
\begin{tikzpicture}
\node[scale=0.9]{
\begin{quantikz}[row sep=0.3cm, column sep=0.4cm]
    & \octrl{1} & \octrl{2} & \gate{X\bigl((m_2 + m_3 - m_4) / 2\bigr)} & \gate{S(2)} & \qw \\
    & \targ{} & \qw  & \measuretab{q = m_2-m_4} \\
    & \octrl{1} & \targ{}  & \measuretab{q=m_3 - m_4} \\
    & \targ{} & \qw & \measuretab{q = m_2+m_4} \\
\end{quantikz}.
};
\end{tikzpicture}
\end{equation}

\noindent We can transition the bottom-most CX${}^{\dag}$ gate so that it acts just before $q$-measurements on these modes  using the relation
\begin{align}
    e^{iq_1 p_3}e^{i q_3 p_4} = e^{iq_3 p_4} e^{i q_1 (p_3 + p_4)}.
\end{align}
The resulting CX${}^{\dag}$ between the third and fourth modes just before $q$-measurements can be identified with reinterpretation of the $q$-measurement of the fourth mode by 
\begin{align}
  m_2+m_4 \mapsto (m_2+m_4)+(m_3-m_4)
\end{align}
using 
\begin{align}
   \bra{m_j}_{q_j}\bra{m_k}_{q_k}C\!X_{ jk}^{\dagger}=\bra{m_j}_{q_j}\bra{m_j+m_k}_{q_k}. 
\end{align}
 As a result, the circuit~\eqref{eq:four_beamsplitters} is equivalent to the following circuit.
\begin{equation}
\begin{tikzpicture}
    \node[scale=0.9]{
\begin{quantikz}[row sep=0.3cm, column sep=0.4cm]
    & \octrl{1} & \octrl{2} & \octrl{3} & \gate{X\bigl((m_2 + m_3 - m_4) / 2\bigr)} & \gate{S(2)}  & \qw \\
    & \targ{} & \qw & \qw & \measuretab{q = m_2-m_4} \\
    & \qw & \targ{} & \qw  & \measuretab{q=m_3 - m_4} \\
    & \qw & \qw & \targ{}   & \measuretab{q = m_2+m_3} \\ 
\end{quantikz}
    };
\end{tikzpicture}
\label{eq:four_beamsplitter_identified}
\end{equation}
This, as well as the circuit~\eqref{eq:beamsplitter_cz}, completes the reduction from the circuit in FIG.~\ref{fig:circuit_identity}\,A to the one in FIG.~\ref{fig:circuit_identity}\,B. We see that this version of the circuit is highly symmetric, as the order of the CX gates can be interchanged (since they commute). 

Now that we have simplified the measurement circuit, we show how the CZ gates associated with each entangled pair can be ``pushed through'' from the satellite modes to the central modes. 

\subsection{Commuting CZ gates to the measurement}
Recall that the circuit in FIG.~\ref{fig:circuit_identity}\,B is given by the following:
\begin{equation}
\begin{tikzpicture}
    \node[scale=0.8]{
\begin{quantikz}[row sep=0.3cm, column sep=0.4cm]
    &\ctrl{4} & \qw & \qw & \qw
    & \octrl{1} & \octrl{2} & \octrl{3} & \gate{X\bigl((m_2 + m_3 - m_4) / 2\bigr)} & \gate{S(2)}  & \qw \\
    &\qw & \ctrl{4} &\qw &\qw 
    & \targ{} & \qw & \qw & \measuretab{q = m_2-m_4} \\
    &\qw & \qw & \ctrl{4} & \qw 
    & \qw & \targ{} & \qw  & \measuretab{q=m_3 - m_4} \\
    &\qw & \qw & \qw & \ctrl{4} 
    & \qw & \qw & \targ{}   & \measuretab{q = m_2+m_3} \\
    \qw & \control{} & \qw & \qw & \qw & \qw & \qw \\
    \qw & \qw & \control{} & \qw & \qw & \qw & \qw  \\
    \qw & \qw & \qw & \control{} & \qw & \qw & \qw \\
    \qw & \qw & \qw & \qw & \control{} & \qw & \qw  
\end{quantikz},
    };
\end{tikzpicture}
\label{eq:circuit_fig_2_b}
\end{equation}
where the circuit input is either $\ket{\bar{+}}$ or $\ket{0}_p$ state (and additionally $\ket{+\bar{T}}$ for the central mode).

Now we conjugate all the CZ gates across the CX${}^{\dag}$ gates using the relations 
\begin{align}
    e^{i q_i p_j}e^{i q_j q_k} = e^{i(q_i + q_j) q_k} e^{i q_i p_j}.
\end{align}
This generates three extra CZ gates, all with support on the central mode. The CZ gates with support on satellite modes precede $q$ homodyne measurements, and these can be omitted by using 
\begin{align}
    \bra{m}_{q_j}C\!Z_{ jk}=\bra{m}_{q_j}Z_k(m).
\end{align}
Following these steps, the circuit~\eqref{eq:circuit_fig_2_b} is made equivalent to the following:

\begin{equation}
\begin{tikzpicture}
    \node[scale=0.8]{
\begin{quantikz}[row sep=0.25cm, column sep=0.3cm]
  & \octrl{1} & \octrl{2} & \octrl{3} & \ctrl{4} & \ctrl{5} & \ctrl{6} & \ctrl{7}  & \gate{X\bigl((m_2 + m_3 - m_4) / 2\bigr)} & \gate{S(2)} & \qw \\
    &\targ{} & \qw & \qw &\qw &\qw & \qw & \qw  & \measuretab{q = m_2-m_4} \\
    & \qw & \targ{} & \qw & \qw & \qw & \qw & \qw   & \measuretab{q=m_3 - m_4} \\
    & \qw & \qw & \targ{} & \qw & \qw & \qw & \qw   & \measuretab{q = m_2+m_3} \\
    &\qw & \qw & \qw & \control{} & \qw & \qw & \qw  & \qw & \qw \\
    &\qw & \qw & \qw & \qw & \control{} & \qw & \qw  & \gate{Z(m_2-m_4)} & \qw  \\
    &\qw & \qw & \qw & \qw & \qw & \control{} & \qw & \gate{Z(m_3-m_4)} & \qw  \\
    & \qw & \qw & \qw & \qw & \qw & \qw & \control{} & \gate{Z(m_2+m_3)} & \qw
\end{quantikz}
    };
\end{tikzpicture}
\label{eq:identity_one_macronode}
\end{equation}

Since outcome-dependent displacement operators $Z(m)$ in the sixth through eighth modes in the circuit~\eqref{eq:identity_one_macronode} commute with CZ gates, we can push them to the circuit inputs.  Furthermore, since the circuit identities \eqref{eq:replacing_beamsplitter}--\eqref{eq:identity_one_macronode} do not use the properties of input states, we can carry out these identities for all the macronodes at the same time without conflict.  By performing circuit identities for all the macronodes, therefore, we have the following circuit structure for each macronode:
\begin{equation}
\begin{tikzpicture}
    \node[scale=0.8]{
\begin{quantikz}[row sep=0.25cm, column sep=0.3cm] 
    & \gate{Z_1} & \octrl{1} & \octrl{2} & \octrl{3} & \qw & \ctrl{4} & \ctrl{4} & \ctrl{4} & \ctrl{4}  & \gate{X_{0}} & \gate{S(2)} & \qw \\
    & \gate{Z_2} &\targ{} & \qw & \qw & \measuretab{q = m_2-m_4} \\
    & \gate{Z_3} & \qw & \targ{} & \qw & \measuretab{q=m_3 - m_4} \\
    & \gate{Z_4} & \qw & \qw & \targ{}  & \measuretab{q = m_2+m_3} \\
    & & & & & & & & & \\
    & & & & & & & \vdots
\end{quantikz}
    \label{eq:identified_circuit}};
\end{tikzpicture},
\end{equation}
where $Z_j$ denotes byproduct displacements that are pushed into the circuit inputs when we carry out the circuit identities (see the circuit~\eqref{eq:identity_one_macronode}).
In the above circuit, $X_{0}$ depends on homodyne outcomes of satellite modes in this macronode while each $Z_{j}$ $(j\in\{1,2,3,4\})$ depends respectively on outcomes of satellite modes in the neighboring macronode-$i$ (see FIG.~\ref{fig:circuit_identity}).  

By using the commutation relation between $Z_j$ and a CX${}^{\dag}$, i.e. $C\!X_{ jk}^{\dag} Z_j(a) Z_k(b)  = Z_j(a+b) Z_k(b)\, C\!X_{ jk}^{\dag}$, as well as the fact that $Z_j$ does not affect $q$-measurement, the circuit~\eqref{eq:identified_circuit} can further be identified with the following.

\begin{equation}\label{eq:final}
\begin{tikzpicture}
    \node[scale=0.75]{
\begin{quantikz}[row sep=0.25cm, column sep=0.3cm] 
    & \octrl{1} & \octrl{2} & \octrl{3} & \qw & \ctrl{4} & \ctrl{4} & \ctrl{4} & \ctrl{4}  & \gate{X_{0}Z_{1}Z_{2}Z_{3}Z_{4}} & \gate{S(2)} & \qw \\
    &\targ{} & \qw & \qw & \measuretab{q = m_2-m_4} \\
    & \qw & \targ{} & \qw & \measuretab{q=m_3 - m_4} \\
    & \qw & \qw & \targ{}  & \measuretab{q = m_2+m_3} \\
    & & & & & & & & & \\
    & & & & & & \vdots
\end{quantikz}
    };
\end{tikzpicture} 
\end{equation}
This completes the reduction from the circuit in FIG.~\ref{fig:circuit_identity}\,B to that in FIG.~\ref{fig:circuit_identity}\,C.

Note that the above circuit involves displacements that depend on the outcomes of the three $q$ measurements on the satellite modes. If the measurement outcomes are known precisely, then these byproduct displacements can be undone in classical post-processing: adding or subtracting from the measured values of central modes. In the presence of noise (such as loss or finite squeezing effects), the true outcomes will be displaced by an unknown random amount. When byproducts are corrected, noise in the measurement outcomes will propagate onto the central modes, which will be analyzed in Appendix~\ref{subsec:error_analysis}.

\subsection{Decoupling satellite modes from central mode}
If there is at least a single GKP state ($\ket{\bar{+}}$ or $\ket{+\bar{H}}$) present in a given macronode, then we assume that the top mode of the four-mode circuits considered in this appendix are input with a GKP state. In reality, whether the top mode is GKP or a squeezed state is determined at random at every time step by the source. However, we note that permutations on the inputs of each four-body measurement can be taken into account by adapting the homodyne bases~\cite{alexander2016flexible}, which boils down to changing which mode is measured in $\hat{p}$. We will not explicitly consider all configurations here. Rather, we proceed as if the first mode was a GKP state (unless all four sources emitted squeezed states), and assume the homodyne measurements will be adapted to compensate. In this setting, the first three CX${}^{\dagger}$ gates in the circuit~\eqref{eq:final}, i.e.,
\begin{equation}
\begin{quantikz}[row sep=0.3cm]
    & \octrl{1} & \octrl{2} & \octrl{3} & \qw & \qw \\
    &\targ{} & \qw & \qw & \measuretab{q = m_2-m_4}\\
    & \qw & \targ{} & \qw & \measuretab{q=m_3 - m_4} \\
    & \qw & \qw & \targ{} & \measuretab{q = m_2+m_3}
\end{quantikz}
\label{eq:sum_gate_network}
\end{equation}
act trivially on the circuit inputs.  This is because
\begin{align}
    C\!X_{jk}^{\dagger}\ket{\bar{\psi}}_{j}\ket{\bar{+}}_{k}=\ket{\bar{\psi}}_{j}\ket{\bar{+}}_{k}
\end{align}
and
\begin{align}
    C\!X_{jk}^{\dagger}\ket{\varphi}_{j}\ket{0}_{p_k}=\ket{\varphi}_{j}\ket{0}_{p_k}
\end{align}
hold, where $\ket{\varphi}$ can be an arbitrary state. Therefore, satellite modes in each macronode are decoupled from the central mode. Note that the residual state supported over the central modes of each macronode still depends on the measurement outcomes of each satellite mode. If a satellite mode is in the GKP $\ket{\bar{+}}$ state, then the corresponding byproduct displacement is a shift by multiple of $\sqrt{\pi}$ depending on the measurement outcome.  If it is in the $\ket{0}_p$ state, then the corresponding byproduct displacement is a completely random displacement depending again on the measurement outcome.  When states are not ideal, the byproduct displacements that depend on measurement outcomes would be sampled from a noisy version of the ideal distribution, and will be analyzed later.

To sum up, by applying a specific choice of homodyne measurement bases, the resource state prepared by FIG.~\ref{fig:layer} can be made equivalent to a canonical RHG lattice supported only over the central modes, up to byproduct displacements depending on $q$-homodyne outcomes of the satellite modes (in the same macronode and the nearest-neighbor macronodes) plus a squeezing by a factor of $2$ on each central mode. 
It is not surprising that even in the ideal case there are byproduct displacement operators to keep track of in the computation. This is very similar to computation using CV cluster states or even the teleportation primitive, where there are explicit measurement-dependent displacement operators to keep track of. 

In the next section we explicitly describe how a $p$- or $q$-homodyne measurement outcome on a canonical CV cluster state can be simulated on the macronode lattice state.  Note that $p$- and $q$-homodyne measurements as well as magic state injection are the minimal requirements for universal quantum computation with a canonical RHG cluster state \cite{raussendorf2006}.

\subsection{Macronode to canonical cluster state dictionary}
\label{subsec:processing}
Here we consider how to simulate a measurement of either $\hat{q}$ or $\hat{p}$ on the canonical lattice (where each site has only a single mode) by an appropriate choice of measurements on the macronode lattice. 

\begin{figure}
    \centering
    \includegraphics[width=\columnwidth]{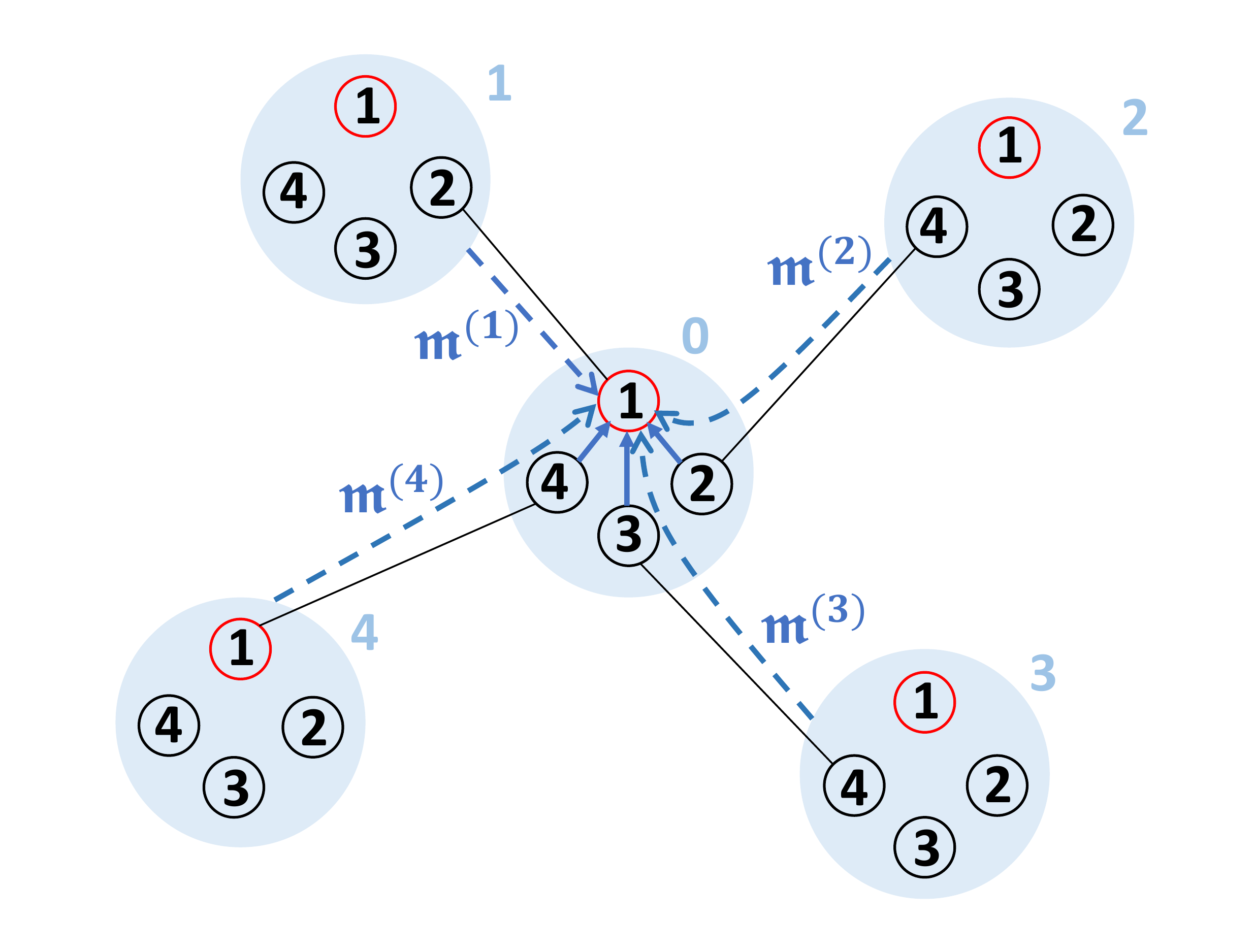}
    \caption{Schematic of the dependency of byproduct displacements.  A group of circles colored in cyan denotes a macronode.  A red circle denotes a central mode and a black circle denotes a satellite mode.  Modes connected by thin black lines are the initial two-mode cluster states that form the lattice.  The solid blue arrow shows the dependency of byproduct displacements in the $q$-quadrature, and a broken blue arrow shows the dependency of byproduct displacements in the $p$-quadrature (see circuit~\eqref{eq:identity_one_macronode} and \eqref{eq:final}).}
    \label{fig:my_label}
\end{figure}

We consider a macronode, labelled 0, and its four neighboring macronodes, enumerated 1--4 (see FIG.~\ref{fig:circuit_identity}).  In each macronode, we proceed under the assumption that the central mode is the top mode, which we enumerate with a 1, and the satellite modes are numbered from 2 to 4.  All the homodyne outcomes are accompanied by a superscript, denoting the macronode index, and a subscript, denoting the mode index. For example, $m_{1,p}^{(0)}$ denotes the $p$-homodyne outcome of the central mode in the macronode-0.  Without loss of generality, we further assume that $i$-th mode of the macronode-0 is connected to a mode in the macronode-$i$.  (See FIG.~\ref{fig:my_label}.)

First, we consider the case in which the central mode in a macronode is measured in $\hat{q}$ in order to simulate the $q$-homodyne outcome $m_{{\rm can}, q}^{(0)}$ of the canonical RHG lattice.  Letting $m_{1,q}^{(0)}$ be the $q$-homodyne outcome of the central mode, we can push the squeezing operator $S(2)$ in the circuit~\eqref{eq:final} into the $q$-homodyne measurement, which results in rescaling the outcome to be $ m_{1,q}^{(0)}/2$.  Furthermore, the byproduct operator $X_0$ in the circuit~\eqref{eq:final} is explicitly given in Eq.~\eqref{eq:identity_one_macronode} by $X_0=X\bigl((m_2^{(0)} + m_3^{(0)} - m_4^{(0)}) / 2\bigr)$.  By pushing this byproduct displacement into the $q$-homodyne measurement in $0$th macronode, we can simulate the $q$-homodyne outcome $m^{(0)}_{\mathrm{can},q}$ on the canonical RHG lattice by the following formula:
\begin{equation}
    \bigl(\hat{q}^{(0)}_1 = \bigr) m^{(0)}_{\mathrm{can},q}   = \frac{m_{1,q}^{(0)} - (m_2^{(0)} + m_3^{(0)} - m_4^{(0)})}{2},
    \label{eq:can_q}
\end{equation}
where $\hat{q}_1^{(0)}$ denotes the quadrature operator of the input of the central mode. In other words, we subtract the value $(m_2^{(0)} + m_3^{(0)} - m_4^{(0)})$ from the result of the position measurement. 

Next, we consider the case in which the central mode is measured in $\hat{p}$ in order to simulate the $p$-homodyne outcome $m_{{\rm can}, p}^{(0)}$ of the canonical RHG lattice.  Letting $m_{1,p}^{(0)}$ be the $p$-homodyne outcome of the central mode, we can push squeezing operator $S(2)$ in the circuit in Eq.~\eqref{eq:final} to $p$-homodyne measurement, which results in rescaling the outcome to be $2 m_{1,p}^{(0)}$.
The circuit in Eq.~\eqref{eq:final} also involves four combined displacements in momentum $Z_1 Z_2 Z_3 Z_4$ that act on the central mode. These displacements depend on particular linear combinations of the measurement outcomes of satellite modes on macronodes 1--4, which we now describe. We denote $Z_i = Z(\mathfrak{m}^{(i)})$, where $\mathfrak{m}^{(i)}$ is given by
\begin{itemize}
    \item $\mathfrak{m}^{(i)} = 0$ (meaning no byproduct displacement) if the $i$-th mode in macronode-0 is connected to the first mode (i.e., the central mode) in macronode-$i$,
    \item $\mathfrak{m}^{(i)} = m^{(i)}_2 - m^{(i)}_4 $ if the $i$-th mode in macronode-0 is connected to the second mode in macronode-$i$,
    \item $\mathfrak{m}^{(i)} = m^{(i)}_3 - m^{(i)}_4 $ if the $i$-th mode in macronode-0 is connected to the third mode in macronode-$i$,
    \item $\mathfrak{m}^{(i)} = m^{(i)}_2 + m^{(i)}_3 $ if the $i$-th mode in macronode-0 is connected to the fourth mode in macronode-$i$.
\end{itemize}
This follows from inspection of the circuit~\eqref{eq:identity_one_macronode} and \eqref{eq:final}.  The fact that the satellite modes are decoupled from the central mode as shown in FIG.~\ref{fig:circuit_identity}\,C has implications on the possible values $\mathfrak{m}^{(i)}$ can take; it is restricted to be an integer multiple of $\sqrt{\pi}$ if the pre-measurement state in the macronode-$i$ is $\ket{\bar{+}}$, and $\mathfrak{m}^{(i)}$ is an unrestricted real number if the pre-measurement state is $\ket{0}_p$. By pushing these byproduct displacements $Z_i$ into the $p$-homodyne measurement of the macronode-0, we can simulate the $p$-homodyne measurement outcome $m^{(0)}_{\mathrm{can},p}$ on the canonical RHG lattice by the following formula:
\begin{equation}
    \biggl(\hat{p}^{(0)}_1 + \sum_{i=1}^{4}\hat{q}^{(i)}_1 = \biggr)m^{(0)}_{\mathrm{can},p}  = 2 m_{1,p}^{(0)} - \sum_{i=1}^{4}\mathfrak{m}^{(i)},
    \label{eq:can_p}
\end{equation}
where $\hat{q}_1^{(i)}$ and $\hat{p}_1^{(i)}$ are quadrature operators of input quantum states of central modes in macronode-$i$ \cite{bourassa2020}. Therefore, we can take these displacements into account by doing subtraction appropriately on the measurement data. 

\subsection{Treatment of the boundary}
In our architecture, there must always be four modes in a macronode.  Therefore, the treatment of the boundary of the cluster state requires extra care.  On the boundary, we always generate additional bipartite entangled states so that the boundary macronode always has four modes, and erase each extra mode, i.e., one of the modes of the additional bipartite entanglement that does not go through beamsplitters, by measuring it in the $\hat{q}$-basis.  In order to eliminate the effect of extra mode, we subtract a measurement outcome of erased mode from $m_{{\rm can}, p}^{(0)}=\hat{p}^{(0)}_1 + \sum_{i=1}^{4}\hat{q}^{(i)}_1$.  In this way, we can use the same passive circuit and circuit identities even on the boundary at the expense of consuming extra bipartite entangled states, which will be negligible when the 3D cluster state is large.

\section{Error analysis}
\label{subsec:error_analysis}

As explained in the main text, in the presence of finite-squeezing and uniform loss, all the measurement outcomes $m^{(i)}_j$ (after rescaling by the factor $1/\sqrt{\eta}$ to transform uniform loss into Gaussian random displacement) acquire Gaussian random noise. We denote these randomly shifted outcomes that we actually obtain in experiment as $m'^{(i)}_j$, i.e.,
\begin{align}
    m^{(i)}_j \xrightarrow{\mathcal{N}[m^{(i)}_j, \epsilon]} m'^{(i)}_j,
\end{align}
where $\mathcal{N}[\mu,\nu]$ denotes normal distribution with mean $\mu$ and variance $\nu$, and $\epsilon=\sigma_{\rm fin.\,sq.}^2 + \frac{1-\eta}{2\eta}$ combines the effect of finite-squeezing and loss.
This also affects the ability to simulate $m^{(i)}_{\rm can}$ of canonical RHG lattice from measurement outcomes of our macronode cluster state.
In this section, we show how errors in the macronode lattice measurements induce errors on the simulated canonical cluster state. Framing the error analysis in terms of the canonical cluster state in this way allows us to then apply the decoding techniques developed in Ref.~\cite{bourassa2020} for the hybrid CV/DV RHG lattice state. 

The analysis in Ref.~\cite{bourassa2020} treated cases where a $\ket{0}_{p}$ state was prepared instead of a $\ket{\bar{+}}$ as a ``swap-out'' error.  In that work, the authors developed a heuristic decoder that assigns an error weight to each mode depending on its measurement outcome. An equivalent decoding strategy can be applied to the present case, with two caveats: (1) the central mode is an encoded $\ket{0}_p$ state only when all the modes in that macronode are in the $\ket{0}_p$ state; and (2) we must take into account additional byproduct operators caused by the satellite modes.
Since we can adopt the same strategy as Ref.~\cite{bourassa2020} if there is a swap-out, we only need to analyse the error weight for the case of no swap-outs in the following.
Hereafter, we call ``performing standard GKP binning for $m'$'' the process of picking the nearest integer multiple of $\sqrt{\pi}$ of $m'$, which is denoted by $\tilde{m}$.

Before going into detail, we give a brief overview of how the noise on measurement data of satellite modes leads to computational errors through byproduct operators. 
For the satellite mode prepared in momentum squeezed states, the best we can do is just to believe the noisy data to cancel the byproduct operators, which leaves a random Gaussian error on the measurement data of the central mode.
The situation is more favourable for satellite modes prepared as GKP states; in the absence of noise, the homodyne outcomes as well as the byproduct displacements from GKP satellite modes must be $0$ mod $\sqrt{\pi}$. Provided that the actual measurement outcome of the satellite mode is not shifted by an amount greater than $\sqrt{\pi}/2$, the noise contribution---that is, the deviation from an integer multiple of $\sqrt{\pi}$---can be corrected by standard GKP binning. However, a larger shift may result in a GKP logical Pauli-$Z$ error on the central mode after the data is processed.  Note finally that we need to rescale the homodyne outcome of the central mode to cancel the squeezing factor in FIG.~\ref{fig:circuit_identity}\,C.  This effectively doubles the noise deviation on the measurement data of the central mode if it is measured in $\hat{p}$-basis while halving it if it is measured in $\hat{q}$-basis. 

To make the description more explicit, we analyze the case of $q$-homodyne $m^{(0)}_{\text{can}, q}$ first. Recall that this is defined via the linear combination of terms given in Eq.~\eqref{eq:can_q}. In the presence of noise, we obtain $m'^{(0)}_{\mathrm{can},q}\coloneqq \frac{m'^{(0)}_{1,q} - (m'^{(0)}_2 + m'^{(0)}_3 - m'^{(0)}_4)}{2}$, where each $m'^{(0)}_j$ is Gaussian randomly shifted from $m^{(0)}_j$ with variance $\epsilon$. Therefore, $m'^{(0)}_{\mathrm{can},q}$ is Gaussian randomly shifted from $m^{(0)}_{\text{can}, q}$ with variance $\epsilon$. We estimate $m^{(0)}_{\text{can}, q}$ by applying standard GKP binning to $m'^{(0)}_{\mathrm{can},q}$, resulting in $\tilde{m}^{(0)}_{\text{can},q}$. The posterior probability of introducing a logical GKP level X error for this measurement (i.e., $\tilde{m}^{(0)}_{\text{can},q}\neq m^{(0)}_{\text{can},q}$) is approximately
\begin{equation}\label{eq:perr1}
    p_{\rm err} = \frac{\sum_{n \in \mathbb{Z}} \exp\bigl[-(m'^{(0)}_{\text{can}, q} - (2n+\bar{r})\sqrt{\pi})^2/(2\epsilon) \bigr]}{\sum_{n \in \mathbb{Z}} \exp\bigl[-(m'^{(0)}_{\text{can}, q} - n\sqrt{\pi})^2/(2\epsilon)\bigr]},
\end{equation}
where $\bar{r} \coloneqq 1 - r \in \{0,1\}$ and $\tilde{m}^{(0)}_{\text{can}, q}/\sqrt{\pi}\equiv r \mod{2}$.

A compact, analytic representation of Eq.~\eqref{eq:perr1} is available in terms of the Jacobi theta function of the third kind, for which we use the notation~\cite{mensen2021}
\begin{align}
    \theta
    (z,\tau)
\coloneqq
    \sum_{n \in \mathbb{Z}}
    \exp
    (
    \pi i n^2 \tau
    +
    2 \pi i n z
    )
    ,
\end{align}
with $z,\tau \in \mathbb{C}$, $\rm{Im}(\tau)>0$. This function is particularly useful for representing a $T$-periodic Gaussian pulse train~\cite{mensen2021}: 
\begin{align}\label{eq:Gauss-train}
    \sum_{n \in \mathbb{Z}}  
    \exp
    \left[
    -\frac
    {(x - n T)^2}
    {2\sigma^2}
    \right]
=
    \sqrt{
    \frac
    {2\pi \sigma^2}
    {T^2}
    }
    \theta
    \left(
    \frac x T, 2\pi i \frac {\sigma^2} {T^2}
    \right)
    ,
\end{align}
Using Eq.~\eqref{eq:Gauss-train}, we can define the auxiliary function
\begin{align}
\label{eq:auxfxn}
    f(x,b,\sigma^2)
&\coloneqq
    \frac{\sum_{n \in \mathbb{Z}} \exp\bigl[-(x - (2n+b)\sqrt{\pi})^2/(2\sigma^2) \bigr]}{\sum_{n \in \mathbb{Z}} \exp\bigl[-(x - n\sqrt{\pi})^2/(2\sigma^2)\bigr]}
\nonumber \\
&=
    \frac
    {
    \theta
    \Bigl(
        \frac {x} {2\sqrt\pi}
        -
        \frac b 2
    ,
        \frac {i \sigma^2} {2}
    \Bigr)
    }
    {
    2
    \theta
    \Bigl(
        \frac {x} {\sqrt\pi}
    ,
        2 i \sigma^2
    \Bigr)
    }
    ,
\end{align}
which will allow us to analytically represent all of the expressions in the rest of this section. For instance, using Eq.~\eqref{eq:auxfxn}, we can rewrite Eq.~\eqref{eq:perr1} as simply 
\begin{align}
    p_{\rm err}
=
    f
    (
    m'^{(0)}_{\text{can}, q}
    ,
    \bar r
    ,
    \epsilon
    )
    .
\end{align}

Now, we analyze the case of $p$-homodyne measurements $m^{(0)}_{\text{can}, p}$ with two extreme examples.  First, we consider the case in which all the modes in macronode-0 are connected to satellite modes in macronode-1 to -4  (i.e., $\mathfrak{m}^{(i)}$ is non-zero), and all these satellite modes are in $\ket{0}_p$ states.  Then, we can have $m'^{(0)}_{\text{can}, p} \coloneqq 2m'^{(0)}_{1,p} - \sum_{i=1}^{4}\mathfrak{m}'^{(i)}$, which is Gaussian randomly shifted from $m^{(0)}_{\text{can},p}$ with variance $12\epsilon$, since $\mathfrak{m}^{(i)}$ is a summation or subtraction of two $q$-homodyne outcomes on satellite modes. As above, we estimate $m^{(0)}_{\text{can},p}$ by applying standard GKP binning to $m'^{(0)}_{\text{can},p}$ to get $\tilde{m}^{(0)}_{\text{can},p}$. The posterior error probability $p_{\rm err}$ for a GKP logical Z error is given by
\begin{align}
    p_{\rm err}
&=
    \frac{\sum_{n \in \mathbb{Z}} \exp\bigl[-(m'^{(0)}_{\text{can}, p} - (2n+\bar{r})\sqrt{\pi})^2/(24\epsilon) \bigr]}{\sum_{n \in \mathbb{Z}} \exp\bigl[-(m'^{(0)}_{\text{can}, p} - n\sqrt{\pi})^2/(24\epsilon)\bigr]},
    \nonumber \\ 
&=
    f
    (
    m'^{(0)}_{\text{can}, p}
    ,
    \bar r
    ,
    12 \epsilon
    )
    ,
\end{align}
where $\bar{r} \coloneqq 1 - r \in \{0,1\}$ and $r \equiv \tilde{m}^{(0)}_{\text{can}, p}/\sqrt{\pi} \mod{2}$, and the auxiliary function~$f$ is defined in Eq.~\eqref{eq:auxfxn}.

Next, we analyze the case in which all the modes in the macronode-0 are connected to satellite modes in macronodes-1 to -4, and all these satellite modes are in $\ket{\bar{+}}$ states.  Then, each $\mathfrak{m}^{(i)}$ must be an integer multiple of $\sqrt{\pi}$ due to the fact that satellite modes are decoupled from the central mode in FIG.~\ref{fig:circuit_identity}\,C.  Therefore, we can estimate $\mathfrak{m}^{(i)}$ by applying standard GKP binning to $\mathfrak{m}'^{(i)}$, resulting in $\tilde{\mathfrak{m}}^{(i)}$. The probability of introducing a logical error by incorrect binning is 
\begin{align}
&\quad
    \frac{\sum_{n \in \mathbb{Z}} \exp[-(\mathfrak{m}'^{(i)} - (2n+\bar{r}^{(i)})\sqrt{\pi})^2/(4\epsilon) ]}{\sum_{n \in \mathbb{Z}} \exp[-(\mathfrak{m}'^{(i)} - n\sqrt{\pi})^2/(4\epsilon)]}
\nonumber \\
&=
    f
    (
    \mathfrak{m}'^{(i)}
    ,
    \bar{r}^{(i)}
    ,
    2\epsilon
    )
    ,
\label{eq:probsatpart}
\end{align}
where $\bar{r}^{(i)}\coloneqq 1 - r^{(i)}  \in \{0,1\}$ and $r^{(i)}\equiv \tilde{\mathfrak{m}}^{(i)}/\sqrt{\pi} \mod{2}$.  From Eq.~\eqref{eq:can_p}, $2 m_{1,p}^{(0)}$ must also be an integer multiple of $\sqrt{\pi}$ in this case.  We can estimate $2 m_{1,p}^{(0)}$ by applying standard GKP binning to $2 m'^{(0)}_{1,p}$, resulting in $2 \tilde{m}^{(0)}_{1,p}$. The probability of introducing a logical error is given by 
\begin{align}
&\quad
    \frac{\sum_{n \in \mathbb{Z}} \exp[-(2 m'^{(0)}_{1,p} - (2n+\bar{r})\sqrt{\pi})^2/(8\epsilon) ]}{\sum_{n \in \mathbb{Z}} \exp[-(2 m'^{(0)}_{1,p} - n\sqrt{\pi})^2/(8\epsilon)]}
\nonumber \\ 
&
= 
    f
    (
    2 m'^{(0)}_{1,p}
    ,
    \bar{r}
    ,
    4\epsilon
    )
    ,
\label{eq:probcentralpart}
\end{align}
where $\bar{r} \coloneqq 1 - r \in \{0,1\}$ and $r \equiv 2 \tilde{m}^{(0)}_{1,p}/\sqrt{\pi} \mod{2}$.  Therefore, we can infer that $\tilde{m}^{(0)}_{\text{can}, p}/\sqrt{\pi} \equiv r - \sum_{i=1}^{4} r^{(i)} \mod{2}$ with a posterior error probability 
\begin{align}
    p_{\rm err}
&\leq
    \frac{\sum_{n \in \mathbb{Z}} \exp[-(2 m'^{(0)}_{1,p} - (2n+\bar{r})\sqrt{\pi})^2/(8\epsilon) ]}{\sum_{n \in \mathbb{Z}} \exp[-(2 m'^{(0)}_{1,p} - n\sqrt{\pi})^2/(8\epsilon)]} \nonumber\\
&\quad
    + \sum_{i=1}^{4} \frac{\sum_{n \in \mathbb{Z}} \exp[-(\mathfrak{m}'^{(i)} - (2n+\bar{r}^{(i)})\sqrt{\pi})^2/(4\epsilon) ]} {\sum_{n \in \mathbb{Z}} \exp[-(\mathfrak{m}'^{(i)} - n\sqrt{\pi})^2/(4\epsilon)]},
    \nonumber \\ 
& = 
    f
    (
    2 m'^{(0)}_{1,p}
    ,
    \bar{r}
    ,
    4\epsilon
    )
+
    \sum_{i=1}^{4} 
    f
    (
    \mathfrak{m}'^{(i)}
    ,
    \bar{r}^{(i)}
    ,
    2\epsilon
    )
    ,
\label{eq:union_bound}
\end{align}
by applying the union bound to Eqs.~\eqref{eq:probsatpart} and \eqref{eq:probcentralpart}.

In generic cases in which some of the connected satellite modes are in $\ket{0}_{p}$ state and others are in $\ket{\bar{+}}$ states, $\mathfrak{m}'^{(i)}$s corresponding to $\ket{\bar{+}}$ satellite modes are binned separately with the variance $2\epsilon$ as explained above, and the rest $2 m'^{(0)}_{1,p} - \sum \mathfrak{m}'^{(j)}$ are binned together with the variance $(4+2t)\epsilon$, where $t$ is the number of $\ket{0}_{p}$ satellite modes in neighboring macronodes. 
The combined posterior error probability $p_{\rm err}$ can be given by the union bound on the same footing as Eq.~\eqref{eq:union_bound}.
Note that, in case the $i$-th mode in the macronode-0 is connected to the central mode in macronode-$i$, no noise is introduced from macronode-$i$ because there is no byproduct (see the circuit~\eqref{eq:identity_one_macronode}). 

\section{Threshold estimation}
\label{sec:app_threshold}

\subsection{Simulation details}
Aided by the preceding sections of the appendix, this section describes how we simulate the error correction of noisy hybrid macronode lattice states and estimate fault-tolerant error thresholds.

First, we generate a hybrid macronode RHG lattice of code distance $d$ and periodic boundary conditions in all three directions. The code distance corresponds to the number of primal unit cells along each dimension, and translates to $4N$ modes in the macronode lattice through $N = 6d^3$. This choice of boundary conditions precludes us having to erase and process superfluous nodes and speeds up the decoding algorithm. By using only the circuit identification of Eq.~\eqref{eq:beamsplitter_cz}, the state generation circuit in FIG.~\ref{fig:layer} is identified as follows:
\begin{equation}
\begin{quantikz}[row sep=0.3cm, column sep=0.5cm]
    \lstick{1} &\ctrl{4} & \qw & \qw & \qw & \qw  & \qw & \qw  & \qw \\
    \lstick{2} &\qw & \ctrl{4} &\qw &\qw & \qw \arrow[u]   & \qw \vqw{1} \arrow[u]  & \qw & \measuretab{q = m_2}  \\
    \lstick{3} &\qw & \qw & \ctrl{4} & \qw & \qw & \qw  & \qw \vqw{1} \arrow[u] & \measuretab{q=m_3}  \\
    \lstick{4} &\qw & \qw & \qw & \ctrl{4} & \qw \arrow[u] & \qw & \qw & \measuretab{q = m_4}   \\
    \lstick{} &  \control{} & \qw & \qw & \qw & \qw & \qw & \qw \\
    \lstick{} &\qw & \control{} & \qw & \qw & \qw & \qw & \qw \\
    \lstick{}  & \qw & \qw & \control{} & \qw & \qw & \qw & \qw \\
    \lstick{} & \qw  & \qw & \qw & \control{} & \qw & \qw  & \qw
\end{quantikz}
\label{circuit:CZBS}
\end{equation}
where each circuit input state is either $\ket{0}_p$ with probability $p_0$ or $\ket{\bar{+}}$ with probability $1-p_0$. We can now permute the mode indices $(1, \ldots, 4N)$ so that central modes are positioned at $\{1+4(i-1) \vert 1 \leq i \leq N \}$ and satellite modes are everywhere else. After the permutation, we generate a list of quadratures of modes at the circuit input $\bm{m}_{\text{in}} = (m^1_q, \ldots m^{4N}_q, m^1_p, \ldots, m^{4N}_p)^T$, where $m^i_p = 0, \ \forall i,$ and
\begin{equation}
m^i_q =
\begin{cases}
\text{rand}(0, \sqrt{\pi}) & \text{if }i\text{-th mode is in } \ket{\bar{+}} \\
\text{randU}(0, 2\sqrt{\pi}) & \text{if }i\text{-th mode is in } \Ket{0}_p
\end{cases}.
\end{equation}
Here the function \text{rand}(a,b) randomly chooses between $a$ and $b$, whereas $\text{randU}(a,b)$ samples from the uniform distribution over the interval $[a, b)$. These quadratures are updated with the application of CZ gates and beamsplitters in Eq.~\eqref{circuit:CZBS} by
\begin{equation}
    \bm{m}_{\text{in}} \to \bm{m}_{\text{out}} = \bm{S}_{BS}\bm{S}_{CZ}\bm{m}_{\text{in}},
\end{equation}
where $\bm{S}_{CZ}$ and $\bm{S}_{BS}$ denotes the symplectic matrices \cite{weedbrook2012} corresponding to CZ gates and beamsplitters.

At this juncture, we generate the noisy homodyne outcomes $m'^i$ with the model $m'^i = \text{randG}(m^i_\text{out}, \epsilon)$, where $\text{randG}(\mu, \nu)$ selects a random sample from the normal distribution $\mathcal{N}[\mu, \nu]$%
\footnote{
Note that this two-pronged sampling procedure is different from the approach taken in~\cite{bourassa2020}. There, one simulates the noise on top of the homodyne outcomes rather than the outcomes themselves, so that one always samples from a normal distribution centred at 0 with covariances supplied by the noise matrix $\bm{\Sigma_0}$. This noise matrix can be diagonal, leaving the samples to be correlated by an application of $\bm{S}_{CZ}$; alternatively, one can sample directly from a multivariate distribution with covariances given by $\bm{S}_{CZ}\bm{\Sigma_0}\bm{S}^T_{CZ}$.
}.
With the simulated noisy outcomes $m'^i$, we use the processing rules in App.~\ref{subsec:processing} to obtain a list $(m^1_{\text{can}, p},...,m^N_{\text{can}, p})$ of effective $p$-homodyne outcomes for the reduced lattice, along with associated conditional qubit-level phase error probabilities $p_\text{err}$ for each effective $p$-homodyne outcome. Furthermore, we label each reduced node its effective type: ``p'' if all modes in the macronode before the reduction are in momentum squeezed states, ``GKP'' otherwise. (This is simply because a mode prepared in the GKP state is always chosen to be a central mode unless there is no GKP state in that macronode).  With these effective outcomes and types, we can construct a canonical RHG lattice of $N$ nodes. This lattice is equivalent to what was fed into the decoder of~\cite{bourassa2020}, except for the boundary conditions (all periodic in our case) and the polarity of the edges (CZ gates all have weight +1 for us). These differences aside, we can run the decoding and recovery operation on the reduced lattice almost exactly as in Algorithms 4 and 5 and Sec. 6 of Ref.~\cite{bourassa2020}. For completeness, we briefly describe the process here.

A single decoding-recovery-verification step for the lattice goes as follows:
\begin{itemize}
    \item A CV (inner) decoder translates the homodyne outcomes to bit values. Although the authors in~\cite{bourassa2020} develop an advanced CV decoder that navigates the correlated noise, we restrict ourselves to standard GKP binning and feed error-weights into the qubit (outer)
    decoder, which appears to have comparable threshold with only a minor effect on logical error rates~\cite{bourassa2020}. 
    \item All the six-body (GKP) Pauli-$X$ stabilizer elements (hereafter referred to as ``stabilizers'') of unit cubes of the primal lattice are identified. In the error-free case, the sum of the bit values associated with each stabilizer should be $0\!\mod{2}$; in other words, each stabilizer ought to have even parity, or else the stabilizer is said to be \emph{unsatisfied}.
    \item A weight is assigned to each node at the interface of two adjacent stabilizers. We use the same combination of heuristic and analog weight assignments as in~\cite{bourassa2020}. Let $n$ be the number of label-``p'' neighbours of a given reduced node. Then we have:
\begin{equation}
    \text{weight}(n) = - \log
    \begin{cases} 
        2/5 & \mbox{if } n = 4, \\
        1/3 & \mbox{if } n = 3, \\
        1/4 & \mbox{if } n = 2, \\
        p_\text{err} & \mbox{if } n \leq 1.
    \end{cases} 
\end{equation}
    \item A \emph{matching graph} is constructed in the following way (we call its edges ``arcs'' for clarity): each of its vertices corresponds to an unsatisfied stabilizer, and each arc has the weight of the shortest-weight path connecting the stabilizers in the lattice. The weight of a path is the sum of the individual weights, and the shortest paths are found using the Dijkstra algorithm~\cite{Dijkstra1959}. 
    \item The matching graph undergoes minimum-weight perfect matching (MWPM) through an implementation of Edmond's algorithm~\cite{Edmonds1965}. The result is a set of pairs of unsatisfied stabilizers (the matching) which minimizes the net weight, that is, finds the likeliest set of error chains that has caused the observed syndrome.
    \item For each pair in the matching, the recovery operation flips the bit values of all the qubits along the path connecting the pair. At this point, all the stabilizers should be satisfied in the resulting lattice. 
    \item If the net effect of the error and recovery is a logical identity, error correction has succeeded; otherwise, a non-trivial logical operator has been applied, and error correction has failed. This can be discovered by computing the total parity of a correlation surface (a plane of primal qubits) of the lattice, with odd parity indicating failure. As we are using periodic boundaries conditions, we must check planes along $x$, $y$, and $z$, unlike~\cite{bourassa2020}, where only one slice was chosen.
\end{itemize}

The above procedure is repeated for roughly 50,000 trials, and the threshold is estimated using the fitting procedure in Ref.~\cite{Harrington2004}.

\begin{table}
    \begin{tabular}{c | c c c c c c c c c}
    \toprule
    $p \, (\%)$ & 0 & 6 & 12 & 18 & 24 & 30 & 36 & 42 & 48 \\
    \midrule
    \midrule
    $t_p (10^{-3})$ & $4.73$ & $3.35$ & $2.30$ & $1.51$ & $0.94$ & $0.53$ & $0.25$ & $0.087$ & $0.016$ \\
    $a_p (10^2)$ & $5.0$ & $6.0$ & $5.3$ & $8.5$ & $28$ & $80$ & $130$ & $190$ & $12000$ \\
    $\nu_p$ & $1.04$ & $1.02$ & $0.99$ & $1.04$ & $1.03$ & $1.06$ & $1.14$ & $1.04$ & $1.31$ \\
    $\mu_p$ & $1.21$ & $1.21$ & $1.14$ & $1.14$ & $1.28$ & $1.34$ & $1.27$ & $1.23$ & $1.42$ \\
    \bottomrule
    \end{tabular}
    \caption{\label{tab:fit_params} Parameters of Eq.~\eqref{eq:pfail_scaling} numerically determined for various swap-out probabilities $p$.}
\end{table}

\subsection{Scaling of the logical error rate} \label{subsec:failure_rate_scaling}

In order to predict the overheads needed for achieving a given logical failure rate, it is desirable to have a scaling law---how the probability of error $P_\text{fail}$ scales with the linear size of the system $d$.
For the case of the (2D, circuit-based) surface code based on physical qubits and Pauli noise, Ref.~\cite{Watson_2014} studied this scaling in detail. The authors found two limiting cases for which analytical expressions can be written. In the region where the probability of error $\kappa$ is below the threshold $\kappa_\text{thr}$, but where the number of errors is large (which happens for $\kappa \gg \frac{1}{d}$), the logical error rate follows a universal scaling law:
\begin{equation} \label{eq:scaling_toric}
    P_\text{fail} = A e^{-a (\kappa_\text{thr} - \kappa)^\mu d},
\end{equation}
where $A, a$ and $\mu$ are constants that can be found numerically. This expression comes from mapping the problem to the random-bond Ising model~\cite{WANG200331}. 
In the regime of low $\kappa$, this expression does not hold anymore, but the logical failure rate is found to be strictly smaller than in Eq.~\eqref{eq:scaling_toric}.

Motivated by the strong connection between the RHG lattice and the surface code~\cite{raussendorf2006}, we consider the following ansatz. For a threshold value $\epsilon^{\text{t}}_p$ and for a given probability $p$ of swap-outs, we define $t_p = \erfc\bigl(\sqrt{\pi}\bigl/(2\sqrt{2\epsilon^{\text{t}}_p}) \bigr)$, where $\erfc$ is the complementary error function. In the regime of high squeezing, $\erfc\bigl(\frac{\sqrt{\pi}}{2\sqrt{2\epsilon}}\bigr)$ gives, to very good approximation, the probability of having a qubit-level error for the noise model considered.  For $\epsilon$ below threshold, we find numerically that, when sufficiently many qubit-level errors happen, the logical error rate $P_{\text{fail}}$ is well-described by the scale-invariant equation~\cite{WANG200331, Watson_2014}
\begin{equation} \label{eq:pfail_scaling}
    P_{\text{fail}} = 0.143 \times \exp\left[-a_p \left( \left[  t_p - \erfc\left(\frac{\sqrt{\pi}}{2\sqrt{2\epsilon}}\right) \right]  d^{\frac{1}{\nu_p}}\right)^{\mu_{p}}\right],
\end{equation}
where $a_p$, $\nu_p$ and $\mu_p$ are found independently for the various swap-out probabilities. Values found are shown in Table~\ref{tab:fit_params}. In the low $\epsilon$ regime, the logical failure rate is found to be below the predicted values. While we believe an analytical expression can be derived for the low $\epsilon$ regime and specific criteria for the validity of Eq.~\eqref{eq:pfail_scaling}~\cite{Watson_2014}, it is beyond the scope of the present work. Eq.~\eqref{eq:pfail_scaling} should thus be interpreted as an upper bound for the logical failure rate.

\begin{figure}
    \centering
    \includegraphics[width=\linewidth]{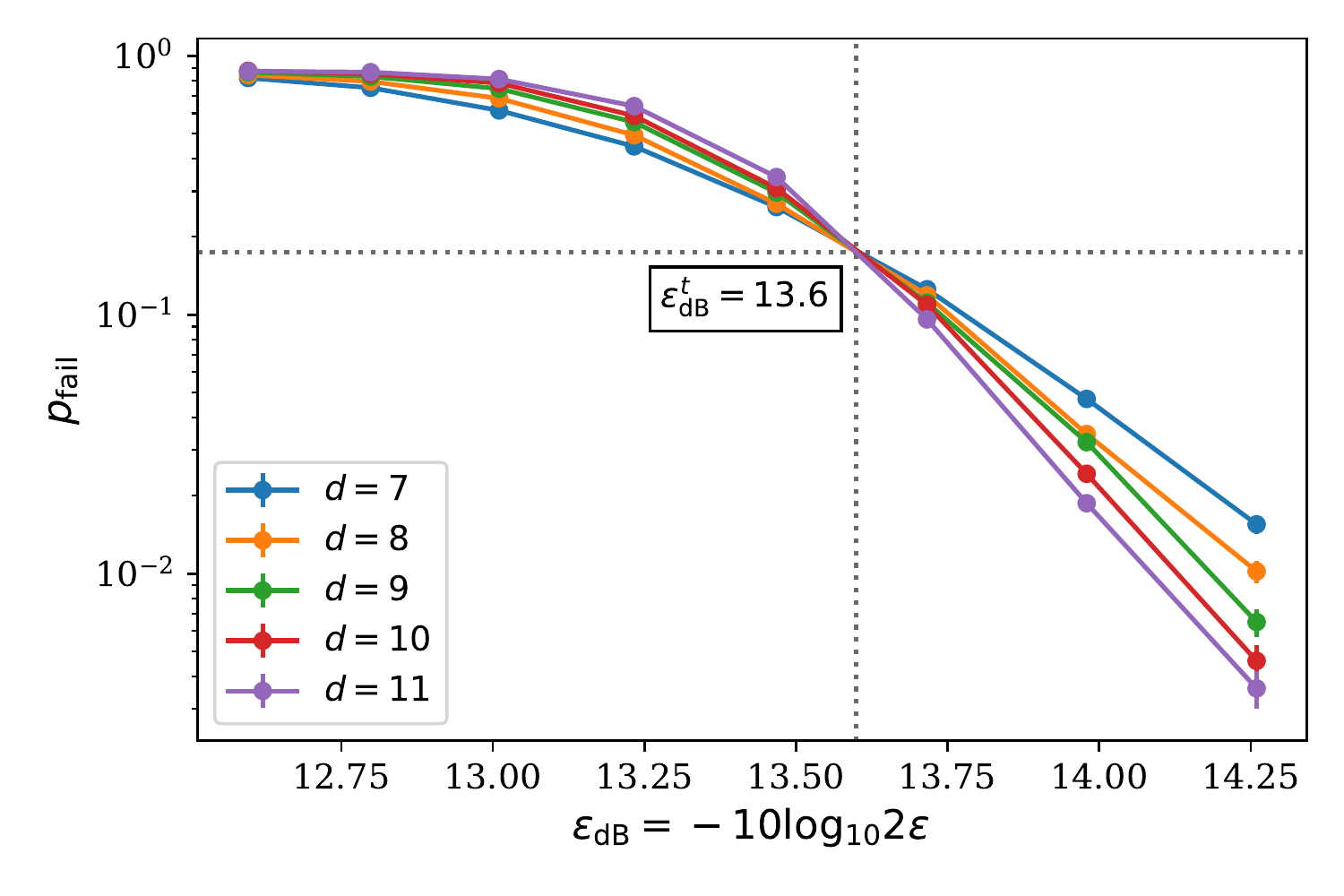}
    \caption{Logical failure probability over noise parameter $\epsilon$ for a macronode RHG code where each macronode is populated by exactly one GKP states and three momentum-squeezed states. Compare with infinite squeezing threshold in FIG.~\ref{fig:swap_tol} at 75\% swap-out rate. With the restriction that each macronode must have a GKP state, no macronode behaves like an effective swapout in the reduced lattice, resulting in a better threshold.
    \label{fig:p_fixed_threshold}
    }
\end{figure}

\subsection{Effect of randomness}

It is instructive to see what would happen to the threshold in the inset of FIG.~\ref{fig:swap_tol} if, instead of demanding every site to be occupied by a GKP state, we require each macronode to have exactly one GKP state. The resulting threshold plot is displayed in FIG.~\ref{fig:p_fixed_threshold}. The threshold (13.6 dB) is worse than in the no swap-out case because the central modes no longer have the benefit of the additional GKP error correction supplied by GKP states in the satellite modes. However, the threshold is non-infinite, an unarguable improvement over the corresponding setting in FIG.~\ref{fig:swap_tol} for 75\% swap-outs. Such is the effect of removing the randomness from state allocation: by demanding a GKP state in each macronode, no mode in the canonical lattice ever behaves like a momentum-squeezed state, yielding no effective swap-outs in the reduced state.

The setting of FIG.~\ref{fig:p_fixed_threshold} is analogous to the no-swap-out case of Ref.~\cite{bourassa2020} with regards to state preparation overheads, since there one also demands every node to contain a GKP state. In that case, the threshold is $\sim 10.5$ dB, ostensibly better than the threshold here. However, recall that the noise from the finitely-squeezed ancillae required for inline squeezing within the CZ gates is not accounted for in~\cite{bourassa2020}. With this taken into account, we expect the threshold to move in the direction of that in FIG.~\ref{fig:p_fixed_threshold}, where the momentum-squeezed states in each macronode also contribute to the noise.

\bibliography{reference.bib}
\end{document}